\documentclass[12pt, a4paper]{article}

\pdfoutput=1

\usepackage{amsmath}
\usepackage{amsfonts}
\usepackage{amssymb}
\usepackage{cite}
\usepackage{multicol,multirow}
\usepackage{graphicx, physics, subcaption}
\usepackage{graphics, epsfig}
\usepackage{epsf}
\usepackage{epstopdf}
\usepackage[usenames,dvipsnames]{color}

\graphicspath{{figs/}}

\usepackage{colortbl}
\definecolor{summersky}{cmyk}{0.71,0.33,0,0.14}
\definecolor{flamingo}{cmyk}{0,0.51,0.71,0.14}
\definecolor{rp}{cmyk}{0.2, 1, 0.6, 0}
\definecolor{pacificblue}{cmyk}{0.95,0.3,0, 0.19}
\definecolor{gray60}{cmyk}{0.4,0.4,0,0.8}

\numberwithin{equation}{section}

\definecolor{ultramarine}{rgb}{0.07, 0.04, 0.56}
\definecolor{cadmiumgreen}{rgb}{0.0, 0.42, 0.24}
\definecolor{indigo(dye)}{rgb}{0.0, 0.25, 0.42}
\usepackage[linktocpage=true,breaklinks]{hyperref}
\hypersetup{
	colorlinks=true,
	citecolor=ultramarine,
	linkcolor=cadmiumgreen,
	urlcolor=indigo(dye),
}

\newcommand{\Mp}{M_{\rm Pl}}
\newcommand{\mR}{\mu}
\newcommand{\mr}{\alpha}
\newcommand{\mn}{\beta}

\def\bea{\begin{eqnarray}}
	\def\eea{\end{eqnarray}}
\def\be{\begin{equation}}
	\def\ee{\end{equation}}
\def\ba{\begin{array}}
	\def\ea{\end{array}}
\usepackage{ulem}
\newcommand\redsout{\bgroup\markoverwith{\textcolor{red}{\rule[0.4ex]{2pt}{2pt}}}\ULon}

\begin{document}
	
\begin{flushright} {\footnotesize YITP-21-119 \\ IPMU21-0065}  \end{flushright}
\vspace{5mm}

\def\thefootnote{\fnsymbol{footnote}}
%%%%%%%%%%%%%%%%%%%%%%%%%%%%%%%%%%%%%%%%%%%%%%%%%%%%%

\begin{center}
{\bf\large Dark Matter from Entropy Perturbations in Curved Field Space}
\\[0.5cm]

{Hassan Firouzjahi$^{1}$, Mohammad Ali Gorji$^{2}$, Shinji Mukohyama$^{2,3}$, Alireza Talebian$^{1}$ }
\\[.7cm]

{\small \textit{$^{1}$School of Astronomy, 
		Institute for Research in Fundamental Sciences (IPM) \\ 
		P.~O.~Box 19395-5531, Tehran, Iran }} \\

{\small \textit{$^{2}$Center for Gravitational Physics,
		Yukawa Institute for Theoretical Physics \\
		Kyoto University, 606-8502, Kyoto, Japan}} \\

{\small \textit{$^{3}$Kavli Institute for the Physics and Mathematics of the Universe (WPI), 
		The University of Tokyo Institutes for Advanced Study, 
		The University of Tokyo, Kashiwa, Chiba 277-8583, Japan}} \\

\end{center}
\vspace{.3cm}
\hrule
%\vspace{0.0cm}
%%%%%%%%%%%%%%%%%%%%%%%%%%%%%%%%%%%%%%%%%%%%%%%%
\begin{abstract}
The accumulated energy density of the excited entropy modes in multiple field inflationary scenarios can play the role of dark matter. In the usual case of a flat field space without any turning trajectory, only light superhorizon entropy modes can be excited through the gravitational instability. In the case of a negatively curved field space, we show that subhorizon entropy modes can be excited as well through the tachyonic instability induced by the negative curvature of the field space. The latter, which is known as the geometrical destabilization mechanism, allows for the production of entropy modes with masses  larger than or at the  order of the Hubble expansion rate during inflation, leading to a new dark matter scenario. Due to the contribution of subhorizon modes, the corresponding spectral density has a peak at a scale smaller than its counterpart in the models based on a flat field space. This difference makes our model observationally distinguishable.
\end{abstract}
\vspace{0.3cm}
\hrule
\newpage
%%%%%%%%%%%%%%%%%%%%%%%%%%%%%%%%%%%%%%%%%%%%%%%%
%\tableofcontents
%%%%%%%%%%%%%%%%%%%%%%%%%%%%%%%%%%%%%%%%%%%%%%%%%%%%%

%\newpage

\section{Introduction}
\setcounter{equation}{0}

Although dark matter constitutes most of the matter content of the universe, its nature is still unknown. This is one of the most compelling evidences which signal physics beyond the Standard Model (SM) of particle physics. There are many dark matter candidates based on physics beyond the SM. Among them the weakly interacting massive particles (WIMPs) are the most well-known scenarios. The WIMPs are mostly expected to be thermally produced during the radiation dominated era more or less in the same way as the production of the SM particles in the standard big bang cosmology. In the absence of detection of WIMPs, the researchers tend to look for the other possibilities like axion dark matter \cite{Marsh:2015xka,Agrawal:2017eqm,Agrawal:2018vin,Machado:2018nqk} and vector dark matter models \cite{Nelson:2011sf,Arias:2012az,Essig:2013lka,Dror:2018pdh,Fabbrichesi:2020wbt,Salehian:2020dsf,Fedderke:2021aqo}. While most of these models deal with the production of dark matter during the radiation dominated era, it is noticed that dark matter particles can be produced at much earlier time during the inflationary stage \cite{Polarski:1994rz, Chung:1998zb, Chung:1998ua, Chung:2018ayg,  Alonso-Alvarez:2018tus,Markkanen:2018gcw,Padilla:2019fju,Tenkanen:2019aij,Herring:2019hbe,Herring:2020cah,Cosme:2020nac,Ling:2021zlj,Graham:2015rva,Bastero-Gil:2018uel,Nakayama:2019rhg,Nakayama:2020rka,Nakai:2020cfw,Kolb:2020fwh,Salehian:2020asa,Firouzjahi:2020whk,Bastero-Gil:2021wsf}. Inflation is an integral part of the standard model of cosmology which solves the horizon and flatness problems and, more interestingly, provides the seed for the  observable structures in the universe. It is then natural to consider the possibility that not only the seeds of the structures in the universe but also the seed of dark matter are produced during the inflationary stage.

Although the inflationary scenario is quite successful in providing an initial condition for the standard big bang cosmology, the origin of the inflaton field, which drives inflation, is not clear yet. Indeed, there is no a priori reason to believe that only one field drives inflation. On the other hand, CMB data are in favour of the single field models as the observed perturbations are adiabatic and no entropy perturbations are detected. This fact may suggest that there is no need for more than one field as the extra fields tend to produce entropy/isocurvature perturbations. However, the accumulated energy density of the excited isocurvature modes can be a dark matter candidate. The source of the isocurvature modes can be, e.g., scalar or vector fields. In the case of vector isocurvature modes, the dark photons are the most well-known scenarios usually dubbed as vector dark matter \cite{Graham:2015rva,Bastero-Gil:2018uel,Nakayama:2019rhg,Nakayama:2020rka,Nakai:2020cfw,Kolb:2020fwh,Salehian:2020asa,Firouzjahi:2020whk,Bastero-Gil:2021wsf}. In these scenarios, nearly massless vector modes can be excited during inflation through interactions with the inflaton or other isocurvature fields which break the conformal symmetry of the vector sector. As a dark matter candidate, after production, the vector field should acquire mass before the time of matter and radiation equality. There are two possibilities: i) the vector field is completely massless during inflation and it acquires mass sometime later through the Higgs symmetry breaking mechanism, ii) the vector field has a mass much less than the Hubble parameter during inflation. In the first case, heavy vector dark matter particles can be produced since the symmetry breaking can happen even right after the time when enough particle production is achieved. In the second case, however, lighter vector dark matter particles can be produced since the vector field becomes massive when the Hubble parameter drops below its mass through the expansion of the universe.

The possibility  of dark matter from scalar isocurvature modes based on inflationary scenarios with multiple scalar fields are vastly studied in the recent years \cite{Polarski:1994rz,Alonso-Alvarez:2018tus,Markkanen:2018gcw,Padilla:2019fju,Tenkanen:2019aij,Herring:2019hbe,Herring:2020cah,Cosme:2020nac,Ling:2021zlj}. As the scalar-tensor theories are not conformally invariant in general, the scalar isocurvature modes can be excited even through their universal interaction with gravity and without any direct interactions with the inflaton field. In this case, only superhorizon modes will be excited. These scalar isocurvature dark matter scenarios only include superhorizon modes in their spectrum. Moreover, these excited isocurvature modes should be almost massless during inflation in order to allow for the efficient gravitational particle production. These are the common features of these types of models which make it hard to distinguish them from each other by the observations.

The inflationary models with curved field spaces are the most general multiple field scenarios that one can consider in the context of scalar-tensor theories without higher derivative terms and without nonlinear kinetic terms. These models have been widely studied in the literature during the recent years. They can provide large non-Gaussianities with different shapes for the curvature perturbations \cite{Seery:2005gb,Gao:2008dt,Elliston:2012ab,Kaiser:2012ak,Tada:2016pmk, Mizuno:2017idt, 
Fumagalli:2019noh,Garcia-Saenz:2019njm,Bjorkmo:2019qno,Ferreira:2020qkf}. The semiheavy fields (compared with the Hubble expansion rate during inflation) can be excited in these models which lead to the observable effects on the CMB spectrum \cite{Chen:2009zp,Achucarro:2010da,Cespedes:2012hu,Achucarro:2012sm,Noumi:2012vr, Emami:2013lma, Wang:2019gok}. Moreover, primordial black holes can form as there is a possibility to enhance the power spectrum of curvature perturbations at small scales in these scenarios \cite{Palma:2020ejf}. As we have mentioned above, these models generally provide entropy perturbations. It is well-known that entropy modes can be enhanced through a tachyonic instability induced by negative curvature of the field space which is called geometrical destabilization \cite{Renaux-Petel:2015mga}. While the effects of the curvature of the field space on the spectrum of the curvature perturbations are widely studied \cite{Achucarro:2019pux, Achucarro:2016fby,Achucarro:2017ing,Peterson:2010np, Cremonini:2010ua,Schutz:2013fua}, the effects on the entropy modes as a source of dark matter are not explored yet. In this paper, having in mind that isocurvature modes can be the source of dark matter, we study the excitation of the entropy modes by the curvature of the field space. We show that apart from the well-known gravitational production of the light superhorizon entropy modes, semiheavy subhorizon entropy modes can also be naturally produced through the geometrical destabilization. This provides a new scenario for the isocurvature dark matter which is observationally distinguishable from all other isocurvature dark matter models that were already studied in the literature.

The structure of the paper is as follows. In Section \ref{sec:review}, we review the most general two-field inflationary scenario with linear kinetic terms, where the field space is generically curved, and decompose the perturbations into the curvature and entropy modes. We obtain the power spectra for the curvature and entropy perturbations in the case of a geodesic trajectory in the field space. In Section \ref{sec-DM-density}, we obtain relic density of dark matter produced from the accumulated energy density of the excited entropy modes. In Section \ref{sec-flat}, we consider the case of light entropy dark matter from multiple inflationary models based on the flat field space and  show that only light superhorizon modes can be excited through the gravitational particle production. In Section \ref{sec-curved}, we consider a particular model for the curved field space and  show that subhorizon modes with arbitrary masses can be excited through the geometrical destabilization mechanism. Section \ref{Conclusion} is devoted to the summary and conclusions. Some technical analysis are presented in appendices \ref{app-rho} and \ref{app-Hamiltonian}.

\section{Two-field inflation}\label{sec:review}

In the context of scalar-tensor theories without higher derivative terms and without nonlinear kinetic terms, the most general multiple inflationary scenario is the one with a curved field space, a general potential and a non-minimal coupling of scalar fields to the spacetime curvature in the Jordan frame. However, the effects of the non-minimal coupling can be removed by a conformal transformation of the spacetime metric from the Jordan frame to the Einstein frame \cite{Kaiser:2010ps,Kaiser:2012ak, McDonough:2020gmn}. In the Einstein frame, the most general multiple field inflationary scenario without higher derivative terms and without nonlinear kinetic terms is given by the following action
\begin{equation} \label{total-action}
S = \int {\rm d}^4 x \, 
\sqrt{-g}\left[\frac{\Mp^2}{2} {\mathrm R} 
-\frac{1}{2}g^{\alpha\beta} \gamma_{a b}(\phi^c) \partial_{\alpha}\phi^a\partial_{\beta}\phi^b-V(\phi^a)\right] \,.
\end{equation}
In the gravitational sector, $\Mp=1/\sqrt{8\pi{G}}$ is the reduced Planck mass and $\mathrm{R}$ is the Ricci scalar which is constructed from the spacetime metric $g_{\mu\nu}$. The system also  includes the scalar fields $\phi^a=(\phi^{1},\phi^{2})$ which span the field space with the metric $\gamma_{a b}(\phi^c)$ and also the potential $V(\phi^a)$. For the sake of simplicity, we have restricted our analysis to the case of two fields while the analysis for the case with a larger (but finite) number of fields is quite straightforward thanks to the covariant formalism that we review below \cite{GrootNibbelink:2000vx,GrootNibbelink:2001qt,Langlois:2008mn,Gao:2009at,Achucarro:2010da,Gong:2011uw}. In order for the kinetic energies to be  bounded from below, we assume that the field space metric $\gamma_{a b}(\phi^c)$ is positive definite. 

The Einstein equations can be obtained by taking the variation of the action \eqref{total-action} with respect to the metric as
\begin{eqnarray}\label{EEs}
\Mp^2 G_{\mu\nu} = T_{\mu\nu} \,; \hspace{1cm}
T_{\mu\nu} = \gamma_{a b}\partial_{\mu}\phi^a\partial_{\nu}\phi^b
- \Big( \frac{1}{2}g^{\alpha\beta} \gamma_{a b}\partial_{\alpha}\phi^a\partial_{\beta}\phi^b 
+ V(\phi^a) \Big) g_{\mu\nu} \,,
\end{eqnarray}
where $G_{\mu\nu}$ is the Einstein tensor and $T_{\mu\nu}$ denotes the energy-momentum tensor of the scalar fields. Taking variation with respect to the scalar fields, we find
\begin{eqnarray}\label{KGE}
\Box \phi^a + \Gamma^{a}_{bc} \partial_\alpha \phi^b \partial^{\alpha} \phi^{c} - V^a = 0 \,,
\end{eqnarray}
where $ V^a \equiv \gamma^{ab} V_b$ with $V_{b} = \partial_b V$, 
\begin{equation}\label{christoffel}
\Gamma^{a}_{b c} = \frac{1}{2} \gamma^{ad} (\partial_b \gamma_{dc}  + \partial_c \gamma_{bd} - \partial_d \gamma_{bc} ) \,,
\end{equation}
is the Christoffel symbol in the field space and $\gamma^{ab}$ is the inverse of the field space metric $\gamma_{ab}$. 

The Riemann tensor in the field space is given by
\begin{equation}
\mathbb{R}^{a}{}_{b c d} = \partial_{c} \Gamma^{a}_{b d} - \partial_{d} \Gamma^{a}_{b c} + \Gamma^{a}_{c e} \Gamma^{e}_{d b} -  \Gamma^{a}_{d e} \Gamma^{e}_{c b} \,.
\end{equation}
Since we restrict our analysis to the case of two scalar fields, the Riemann tensor can be written as
\begin{equation}\label{Riemann}
\mathbb{R}_{a b c d} = \frac{1}{2} \mathbb{R} (\gamma_{a c} \gamma_{b d} - \gamma_{a d} \gamma_{c b} ) \,,
\end{equation}
where $\mathbb{R} = \gamma^{a b} \mathbb{R}^{c}{}_{a c b}$ is the field space Ricci scalar.

\subsection{Background analysis}

For the homogeneous and isotropic cosmological background, we consider a spatially flat FLRW geometry together with homogeneous vacuum expectation values for the scalar fields as follows
\begin{equation}\label{metric}
\dd s^2 = -\dd t^2+a^2(t)\delta_{ij} \, \dd x^i \dd x^j \,, \hspace{1cm} \phi^a = \varphi^a(t) \,,
\end{equation}
where $t$ is the cosmic time and $a(t)$ is the scale factor. 

Substituting \eqref{metric} in the Einstein equations \eqref{EEs}, we find the Friedmann equations
\begin{eqnarray}
3 \Mp^2 H^2 &=& \frac{1}{2}\dot \sigma^2+V \,, \label{Friedmann-eq} \\
\Mp^2 \dot{H} &=& -\frac{\dot \sigma^2}{2 } \,, \label{Friedmann-eq-2}
\end{eqnarray}
where a dot denotes derivative with respect to the cosmic time $t$, $H = \dot{a} /a$ is the Hubble expansion rate, and we have also defined
\begin{equation}
\dot \sigma^2 \equiv \gamma_{a b} \dot \varphi^a \dot \varphi^b \,.
\end{equation}
In this view, $\dot \sigma^2/2$ measures the kinetic energy along the background trajectory. 

The scalar field equations \eqref{KGE} for the background configuration \eqref{metric} take the form
\begin{eqnarray}\label{KGE-homogeneous-eqs}
D_t \dot{\varphi}^a+3H\dot{\varphi}^a+V^a=0 \,,
\end{eqnarray}
where we have introduced a covariant time derivative ${D}_{t}$ in curved field space as follows
\begin{equation}
D_t X^a = \dot X^a + \Gamma^{a}_{b c} \dot \varphi^b X^c \,,
\end{equation}
for an arbitrary vector field $X^a(t)$ in the field space.

\begin{figure}[t!]
	\begin{center}
		\includegraphics[scale=0.8]{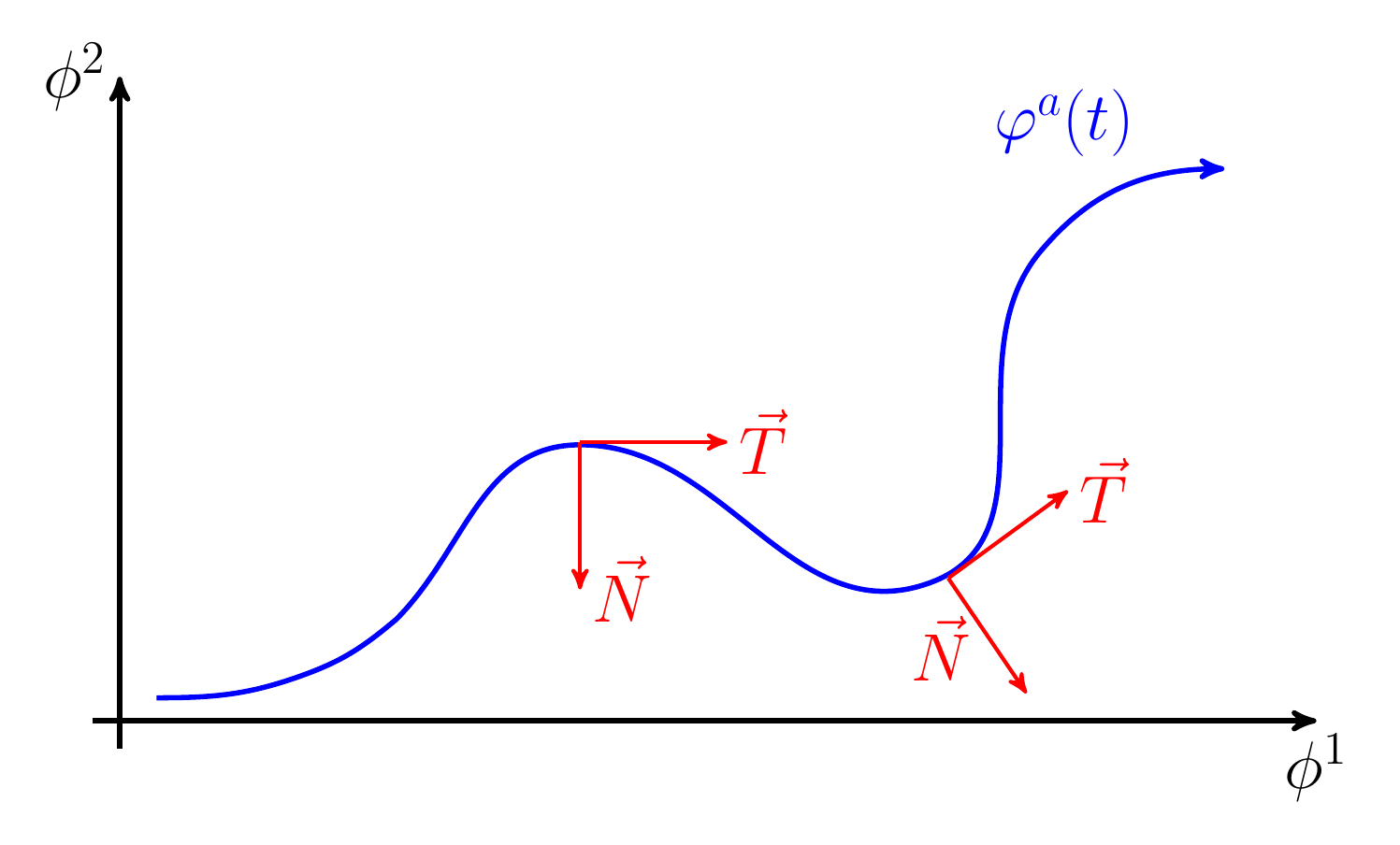}
		\caption{\footnotesize The schematic time evolution of the tangent vector $T^a$ and the normal vector $N^a$ along the background trajectory denoted by the blue curve.  These orthonormal vectors are used to decompose the perturbations $\delta \phi^{a}$ into adiabatic and entropic modes~\cite{Gordon:2000hv}.}
		\label{fig:T-N}
	\end{center}
\end{figure}

Solving background equations \eqref{Friedmann-eq} and \eqref{Friedmann-eq-2} with a proper initial condition, we find a unique solution $\varphi^a(t) = (\varphi^{1}(t) , \varphi^{2}(t))$ which determines the classical trajectory of the system. This trajectory, parametrized by the cosmic time $t$, defines a curve in the field space as shown in Fig. \ref{fig:T-N}. To characterize this curve, it is convenient to consider two unit vectors $T^a$ and $N^a$ as
\begin{eqnarray}\label{T-N}
T^a \equiv \dfrac{\dot \varphi^a}{\dot \sigma} \,, \hspace{1cm}
N^a \equiv \gamma^{ab} \left( \det  \gamma \right)^{1/2} \epsilon_{bc} T^{c} \,,
\end{eqnarray}
where $\epsilon_{a b}$ is the two-dimensional Levi-Civita symbol with $\epsilon_{11} = \epsilon_{22} = 0$ and $\epsilon_{12} = - \epsilon_{21} = 1$. As shown in Fig. \ref{fig:T-N}, the vectors $T^a$ and $N^a$ are respectively tangent and normal to the background trajectory  at any moment. From \eqref{T-N} we see that 
\begin{eqnarray}\label{T-N-norm}
T_a T^a = 1 = N_a N^a \,, \hspace{1cm}  T^a N_a = 0 \,,
\end{eqnarray}
where we have used $\gamma^{ab}$ and $\gamma_{ab}$ to raise and lower the field space indices. 

We can use these two unit vectors to expand any vector field in the field space as $X^a= X_T T^a + X_N N^a$ where $X_T \equiv T_a X^a$ and $X_N \equiv N_a X^a$ are projection along the tangent and normal directions respectively. The equations of motion for the scalar fields (\ref{KGE-homogeneous-eqs}), projected along the tangent direction, gives
\begin{equation}\label{EOM-sigma}
\ddot{\sigma} + 3H\dot \sigma + V_{\sigma} = 0 \,; \hspace{1.5cm} V_{\sigma} \equiv T^a \partial_a V \,,
\end{equation}
while projecting along $N^a$ gives
\begin{equation}\label{time-deriv-T}
D_t T^a=-\frac{V_N}{\dot \sigma}N^a \,; \hspace{1.5cm} 
V_N \equiv N^a \partial_a V \,.
\end{equation}

To characterize the dynamics of the background quantities, we define the following dimensionless parameters
\begin{eqnarray}\label{epsilon}
\epsilon  \equiv  -\frac{\dot{H}}{H^2}  = \dfrac{\dot{\sigma}^2}{2\Mp^2 H^2}\,, 
\hspace{1.5cm}
\eta^a  \equiv - \frac{D_t \dot{\varphi^a}}{H\dot \sigma}.
\end{eqnarray}
Intuitively, $\epsilon$ is similar to the first geometric slow-roll parameter in single field models while $\eta^a$ contains information about the second slow-roll parameter since it includes second time derivatives of the scalar fields. To make this fact more clear, we decompose $\eta^a$ along the tangent and normal directions as
\begin{eqnarray}
\eta^a \equiv \eta_{\parallel} T^a+\eta_{\bot} N^a \,,
\end{eqnarray}
where $\eta_{\parallel} \equiv T_a \eta^a$ and $\eta_{\bot} \equiv N_a \eta^a$ are found to be 
\begin{eqnarray}\label{eta-perp-def}
\eta_{\parallel} = -\frac{\ddot{\sigma}}{H\dot \sigma} \,, \hspace{1.5cm}
\eta_{\bot}  =  \frac{V_N}{H\dot \sigma} \,.
\end{eqnarray}
The above relations make it clear that $\eta_{\parallel} $ is the counterpart of the usual second slow-roll parameter in single field inflation models. Therefore, the slow-roll conditions are given by
\begin{eqnarray}\label{slow-roll}
\epsilon \ll 1 , \hspace{1.5cm} |\eta_{||}| \ll 1 \,. 
\end{eqnarray}
Note that a large value of $\eta_{\bot}$ does not necessarily spoil the above slow-roll conditions. To understand the role of the parameter $\eta_{\bot}$, using (\ref{time-deriv-T}) and (\ref{eta-perp-def}), we find
\begin{eqnarray}\label{geodesics}
D_t T^a = - H \eta_{\bot} N^a \,, \hspace{1.5cm}
D_t N^a = + H \eta_{\bot} T^a \,.
\end{eqnarray}
We see that the vectors $T^a$ and $N^a$ remain covariantly unchanged along the classical trajectory if $\eta_{\bot} = 0$. On the other hand, depending on the sign of $\eta_{\bot}$, the vectors $T^a$ and $N^a$ can turn to the right or to the left during the evolution of the system.

\subsection{Perturbation analysis: curvature/entropy decomposition}

The scalar perturbations around the homogeneous and isotropic background configuration \eqref{metric} are given by
\begin{eqnarray}\label{metric-pert}
\dd s^2= - (1 + 2 A) \dd t^2 + 2 \partial_iB \, \dd t \dd x^i 
+ a^2 \big( ( 1 + 2 \psi ) \delta_{ij} + 2 E_{,ij} \big) \delta_{ij} \, \dd x^i \dd x^j \,,
\end{eqnarray}
and
\begin{eqnarray}\label{field-pert}
\phi^a(t,\textbf{x})=\varphi^a(t)+\delta\phi^a(t,\textbf{x}) \,,
\end{eqnarray}
where $A, B, \psi, E$, and $\delta\phi^a$ are functions of time and spatial coordinates. It is more convenient to work with the gauge-invariant combinations $\delta\phi^a+\frac{\dot{\varphi}^a}{H}\psi$ and, therefore, we parametrize scalar perturbations in terms of curvature perturbation $\mathcal R$ and entropy perturbation $\mathcal S$ as follows 
\begin{eqnarray}\label{curv-isocurv}
\mathcal{R} \equiv
\psi+\frac{H}{\dot \sigma} T_a\delta \phi^a \,, \hspace{1.5cm}
\mathcal{S} \equiv  
\frac{H}{\dot \sigma} N_a \delta \phi^a \equiv   
\frac{H}{\dot \sigma}   {\mathcal F} \,.
\end{eqnarray}

Working in the spatially flat gauge $\psi=0=E$ and integrating out non-dynamical variables $A$ and $B$, it is straightforward to show that the quadratic action for the scalar perturbations in terms of the gauge-invariant quantities ${\cal R}$ and ${\cal F}$ is given by \cite{Cespedes:2012hu}
\begin{eqnarray}
S^{(2)}%_{\rm tot} 
=  \frac{1}{2} \int \! {\rm d}^4 x \, a^3 \bigg[  \frac{\dot \sigma^2}{H^2} \dot {\mathcal R}  ^2  - \frac{\dot \sigma^2}{H^2} \frac{(\nabla \mathcal R)^2}{a^2}  +   \dot {\mathcal F}^2  
- \frac{ (\nabla \mathcal F)^2 }{a^2}
+  4  \eta_{\perp} \dot \sigma \dot{\mathcal R} \mathcal F - m_{\rm eff}^2 \mathcal F^2   \bigg] ,
\quad \label{total-action-R-F}
\end{eqnarray}
where we have defined the effective mass $m_{\rm eff}$  as
\begin{eqnarray}\label{effective-mass}
m_{\rm eff}^2 \equiv m_s^2 + \mR^2 - (H \eta_{\bot})^2 \,;
\hspace{1.5cm}
m_s^2 \equiv V_{NN} \,, 
\hspace{.5cm}
\mR^2 \equiv 
\epsilon \Mp^2 H^2 \, \mathbb{R} = \frac{1}{2} \dot{\sigma}^2 \mathbb{R} \,.
\end{eqnarray}
Here, $V_{NN} \equiv N^a N^b \nabla_a \nabla_b V$ is the conventional mass term. 

It is worth looking at various contributions in $m_{\rm eff}^2$. The first term in $m_{\rm eff}$, as just mentioned, represents the usual contribution. The second term represents the contributions from the curvature of field space while the last term is due to the centripetal force induced by the turn in the field space. In this work we are mainly interested in the effects of the second term, which is responsible for the geometrical destabilization mechanism \cite{Renaux-Petel:2015mga}, while the non-geodesic motion in the field space induced by $\eta_\perp$ has interesting features ~\cite{Palma:2020ejf,Fumagalli:2019noh}. 
%For our purpose in this paper, we focus on the effects coming from the curvature of the field space.

The equations of motion governing the system are given by
\begin{eqnarray}
\ddot {\mathcal R} +  (3  + 2 \epsilon - 2 \eta_{||}) H \dot  {\mathcal R} - \frac{\nabla^2  {\mathcal R} }{a^2} &=& - 2 \frac{H^2}{\dot \sigma} \eta_{\bot} \left[ \dot {\mathcal F} + (3 - \eta_{||} - \xi_{\bot}) H {\mathcal F} \right] , 
\label{eq-of-motion-R}
\\
\ddot {\mathcal F} +  3  H \dot  {\mathcal F} - \frac{\nabla^2  {\mathcal F} }{a^2} + m_{\rm eff}^2  {\mathcal F} &=& 2 \eta_{\bot} \dot \sigma \dot {\mathcal R} , \label{eq-of-motion-F}
\end{eqnarray}
where $\xi_{\bot} \equiv - \dot \eta_{\bot} / (H \eta_{\bot})$. As we see, when $\eta_\perp \neq 0$ the two modes become coupled and it may not be easy to solve them analytically. As we are mostly interested in the effects coming from the curvature of the field space and also  in order to simplify the analysis, in the following we consider  the case where $\eta_\perp=0$, representing geodesic motions in the curved field space.

\subsection{Geodesic trajectory in the field space}\label{sec-PS}

In this section we focus on the geodesic trajectories $\eta_{\bot}=0$ in the field space where the two modes ${\mathcal R}$ and ${\mathcal F}$ 
are decoupled at the linear level.  

In the slow-roll regime \eqref{slow-roll} where we have a quasi-de Sitter expansion, it is better to work with the conformal time $\tau=\int{\rm d}t/a(t)$ in term of which the action Eq.~\eqref{total-action-R-F}
 takes the following form
\begin{eqnarray}\label{conformal-action-R-F}
S^{(2)}%_{\rm tot} 
=  \frac{1}{2} \int \! {\rm d}^3 x \, {\rm d}\tau \, a^2 \bigg[  2\epsilon \left( {\cal R}'^{2}  - (\nabla \mathcal R)^2 \right)  
+   {\cal F}'^{2}  - (\nabla \mathcal F)^2 - \big( m_s^2 + \mR^2 \big) a^2 \mathcal F^2   \bigg] .
\end{eqnarray}

We have assumed that on the background trajectory $\eta_{\bot} = V_N / (H\dot \sigma) =0$, which implies $V_N =0$ but it does not imply $V_{NN}=0$. This is obvious from the fact that the $N^a$ direction is linearly independent of the tangent $T^a$ of the trajectory and therefore demanding that $V_N=0$ everywhere on the trajectory does not restrict the value of $V_{NN}$, i.e. the second derivative of the potential along the independent direction $N^a$. Therefore, $m_s\neq0$ is compatible with $\eta_{\bot}=0$. 

%The equations of motion \eqref{eq-of-motion-R} and \eqref{eq-of-motion-F} decouple as follows
%\begin{eqnarray}\label{eq-of-motion-Geodesic_R} \ddot {\mathcal R} +  (3  + 2 \epsilon - 2 \eta_{||}) H \dot  {\mathcal R} 
%- \frac{\nabla^2  {\mathcal R} }{a^2} &=& 0 ,  \\ \label{eq-of-motion-Geodesic_F} \ddot {\mathcal F} +  3  H \dot  {\mathcal F} 
%- \frac{\nabla^2  {\mathcal F} }{a^2}  + \big( m_s^2 + \mR^2 \big) {\mathcal F} &=& 0 . \end{eqnarray}

To quantize the system and to obtain the power spectra, we define the canonically normalized fields
\begin{align}\label{canonical-fields}
u \equiv z {\cal R} \,,
\hspace{1cm}
{\rm and}
\hspace{1cm}
s \equiv a {\cal F} \,,
\end{align}
where $z^2 \equiv 2\Mp^2 a^2 \epsilon$. In terms of these variables the action takes the canonical form
\begin{align}\label{conformal-action-s-F}
S^{(2)}%_{\rm tot} 
=  \frac{1}{2} \int \! {\rm d}^3 x \, {\rm d}\tau \bigg[  \left( u'^{2}  - (\nabla  u)^2 + \dfrac{z''}{z}u^2 \right)  
+ \left( s'^{2}  - (\nabla  s)^2 + \dfrac{a''}{a}s^2 
- \big( m_s^2 + \mR^2 \big) a^2 s^2 \right)  \bigg] \,.
\end{align}
The canonical fields $u$ and $s$, corresponding to the adiabatic and entropy perturbations, are decoupled at the linear level and, therefore, we study their power spectra separately in the following. 

Going to the Fourier space with the standard definition $X(\tau, {\bf x}) = \int \frac{\dd^3 k}{(2\pi)^3} \, X_{\bf k}(\tau) e^{i {\bf k} \cdot {\bf x}}$ for the scalar field $X$, and then expanding the operator counterpart of the field $u$ in terms of the annihilation and creation operators as usual, the corresponding mode function satisfies the Mukhanov-Sasaki equation
\begin{eqnarray}\label{equ:Mmode}
u_k'' + \left( k^2 - \frac{z''}{z}\right) u_k = 0 \,.
\end{eqnarray}
Imposing the Bunch-Davies (Minkowski) initial condition in the limit $-k\tau \to \infty$, we find
\begin{eqnarray}
u_k =  \dfrac{\sqrt{-\pi \tau}}{2} \, {\rm H}_\nu^{(1)}(x)\, ; 
\hspace{2cm}
x \equiv -k\tau \,,
\hspace{.5cm} 
\mbox{and}
\hspace{.5cm}
\nu \approx \frac{3}{2} + 2 \epsilon - \eta_{||} \,.
\end{eqnarray}
Then the dimensionless power spectrum of the curvature perturbations ${\cal R} = u/ z$ for the superhorizon modes becomes
\begin{align}\label{PS-R}
{\cal P}_{\cal R} = 
\dfrac{k^3}{2\pi^2}  \left.
\vert{\cal R}_k\right\vert_{k\tau \to 0}^2 = \dfrac{H^2}{8\pi^2\Mp^2 \epsilon} x^{3-2\nu} \,.
\end{align}

For the entropy modes, doing the same process of quantization as the curvature perturbations, we find that the corresponding mode function satisfies
\begin{eqnarray}
\label{omega}
s_k'' + \omega_k^2\, s_k = 0 \,; 
\hspace{1.5cm}
\omega_k^2 \equiv k^2 + \frac{\mn}{\tau^2} - \frac{2+\mr}{\tau^2} \,,
\end{eqnarray}
where $\omega_k$ is the frequency and we have defined
\begin{align}\label{alpha-R}
\mr \equiv -\frac{\mR^2}{H^2} 
= -\epsilon \Mp^2 \mathbb{R} \,,
\hspace{1.5cm}
\mn \equiv \frac{m_s^2}{H^2} \,.
\end{align}
In the above relation $\mn$ is a dimensionless parameter that characterizes the usual mass of the entropy modes which is normalized by the Hubble expansion rate. The dimensionless parameter $\mr$ corresponds to the mass induced by the curvature of the field space which can be positive (negative) 
for $\mr>0$ $(\mr<0)$.

Looking at the $\tau^{-2}$ parts of the effective frequency squared $\omega_k^2$, we notice three distinct contributions. The term containing $-2$ represents the usual gravitational particle production in a quasi-de Sitter background. The terms containing $\alpha$ and $\beta$ respectively represent the effects of the field space curvature and the mass. Note that the parameters $\alpha$ and $\beta$ can take either signs and can be time-dependent. We are interested in situations where in regions of the field space 
the combination $\alpha -\beta$ can becomes negative indicating a period of tachyonic instability. However, in order not to destroy the CMB constraints on entropy perturbation the tachyonic growths should be under control and happen only on sub-CMB scales.

Imposing the Minkowski  initial condition on Eq. \eqref{omega} we find
\begin{eqnarray}\label{mode-entropy}
s_k = \dfrac{\sqrt{-\pi \tau}}{2} {\rm H}_\mu^{(1)}(x)\,; \hspace{1.5cm}
\mu \equiv \sqrt{\frac{9}{4} + \mr - \mn} \,.
\end{eqnarray}
Correspondingly,  the dimensionless power spectrum for the entropy perturbations ${\cal S} = (H/\dot{\sigma}) {\cal F} = (H/a\dot{\sigma})s$ is given by
\begin{align}\label{PS-S}
{\cal P}_{\cal S}
= \dfrac{k^3}{2\pi^2} \left.
\vert{\cal S}_k\right\vert_{k\tau \to 0}^2 = \dfrac{H^2}{8\pi^2\Mp^2 \epsilon} x^{3-2\mu} \,.
\end{align}

Now, let us consider the CMB constraints on the parameters of the model. For the curvature perturbations, the scale-dependence of the spectrum is given by the spectral index $n_{\cal R}-1 \equiv 3-2\nu$ as follows
\begin{equation}\label{ns}
n_{\cal R}-1 = - 4 \epsilon + 2 \eta_{||} \,.
\end{equation}
The CMB observations imply $n_{\cal R}-1 = {\cal O}(10^{-2})$ which put a constraint on a combination of the slow-roll parameters as $2\epsilon - \eta_{||} = {\cal O}(10^{-2})$. On the other hand, there is an upper bound on the power spectrum of the superhorizon entropy perturbations ${\cal P}_{\cal S}/{\cal P}_{\cal R} \lesssim 10^{-3}$ 
at three scales $k=\{ 0.002, 0.05, 0.1 \} \, {\rm Mpc}^{-1}$ \cite{Akrami:2018odb}. Using Eqs. \eqref{PS-R} and \eqref{PS-S}, we find that this constraint on the amplitude of entropy perturbations is rewritten as
\begin{align}\label{alpha_bound}
\mr - \mn \lesssim \delta + \frac{1}{9}\delta^2\,, \qquad
 \delta \equiv 6\epsilon - 3\eta_{||} - \frac{9\ln(10)}{2{\cal N}_{\rm CMB}} 
 \simeq 6\epsilon - 3\eta_{||} - 0.17\cdot\frac{60}{{\cal N}_{\rm CMB}}\,,
\end{align}
where ${\cal N}_{\rm CMB}={\cal O}(60)$ is the number of \textit{e}-folds from the horizon exit of the CMB scale to the end of inflation. Thus, the CMB constraints on the spectra of curvature and entropy perturbations imply $\mr - \mn \lesssim - 0.1$. In the case of light modes with $|\mn|\ll1$, the CMB constraint put the direct constraint $\alpha \lesssim -0.1$ on the curvature of the field space. For the case of heavy modes $\beta-\alpha\gg1$, the mode function \eqref{mode-entropy} receives a Boltzmann suppression factor and the entropy modes will not be efficiently excited. Thus, there is no constraint from the CMB scales  on the spectra of the heavy entropy modes and in this case the parameter $\mr$ can acquire any value on CMB scales
as long as $\beta-\alpha\gg1$.

\section{Dark matter relic density}\label{sec-DM-density}

Looking at Eq. \eqref{omega}, we see that $\omega_k^2<0$ for some entropy modes. The negative contributions come from both the gravitational interaction encoded in the term $-2/\tau^2$ and also from the negative curvature of the field space encoded in the term $-\mr/\tau^2$ with $\mr>0$ (${\mathbb R}<0$). Thus, the adiabatic approximation breaks down and the entropy modes with  $\omega_k^2<0$ can be excited.

The accumulated energy density of the excited entropy modes during inflation can play the role of dark matter when they become non-relativistic later due to the expansion of the universe. In appendix \ref{app-rho}, we have computed the energy density of the excited entropy modes during inflation which is given by Eq. \eqref{app-rho-s}. Going to the Fourier space, we find
\begin{eqnarray}\label{rho-s}
\rho_{s,{\rm e}} = \dfrac{1}{2a^4} \int_{k_{\rm min}}^{k_{\rm max}} 
\dfrac{\dd^3 k}{(2\pi)^3}  \, \bigg[ \Big| a\left(\dfrac{s_k}{a}\right)'
\Big|^2  + \left( k^2 + \frac{\mn+\mr}{\tau^2} \right)|s_k|^2 \bigg] \bigg{|}_{\tau=\tau_{\rm e}} \,,
\end{eqnarray}
where the conformal time $\tau_{\rm e}$ represents the time of end of inflation. The integral limits $k_{\rm min}$ and $k_{\rm max}$ are the smallest and largest momenta respectively which can be excited. We determine their explicit values later. Note that we should only consider contributions from the tachyonic modes, say those modes with $k_{\rm min}<k<k_{\rm max}$. Otherwise, one would include contributions from the pure vacuum fluctuations which should be renormalized away. It is also worth mentioning that during inflation, depending on the value of the parameter $\mr$, the Hamiltonian for the entropy modes given by Eq. \eqref{app-Hamiltonian-s} can become negative. This is not an issue for our model as the parameter $\mr$ vanishes 
i.e. $\mr_{\rm r}=0$ for $\dot{\sigma}=0$ when inflaton stops at the end of reheating. Indeed, we need this local tachyonic instability due to the negative curvature of the field space in order to excite dark matter particles. 

It is more convenient to introduce the dimensionless fractional energy density
\begin{eqnarray}\label{Omega-s}
\Omega_{\tiny s,{\rm e}} \equiv  \frac{\rho_{s,{\rm e}}}{3\Mp^2H_{\rm e}^2} \,,
\end{eqnarray}
where $H_{\rm e}$ is the Hubble expansion rate at the end of inflation. The above quantity represents the contribution of the produced entropy particles to the total energy density of the universe at the end of inflation. Having $\Omega_{\tiny s,{\rm e}}$ in hand, we only need to keep track of the evolution of the energy density from the end of inflation until late times through the expansion of the universe. Moreover, in order to keep the model under theoretical control and for simplicity, we consider an instantaneous reheating. Following Ref. \cite{Salehian:2020asa}, in this case the relic density for the dark matter today is given by
\begin{eqnarray}\label{relic-DM}
\Omega_{\tiny s,0} =
\Big(\frac{g_{s*,0}}{g_{s*,{\rm r}}}\Big)
\Big(\frac{\pi^2}{90}g_{*,{\rm r}}\Big)
\Big(\frac{T_0^3T_{\rm r}}{\Mp^2 H_0^2}\Big) \, 
\Big(\frac{a_ {\rm e}}{a_{\rm NR}}\Big) \,
\Omega_{\tiny s,{\rm e}} \,,
\end{eqnarray}
where $T_0$ and $H_0$ are the CMB temperature and the Hubble parameter today, $g_{s*,0}$ and $g_{s*,{\rm r}}$ are the number of relativistic degrees of freedom for the entropy density today and at the time of reheating respectively while $g_{*,{\rm r}}$ is the number of relativistic degrees of freedom for the energy density at the time of reheating. Additionally, $T_{\rm r}$ denotes the reheating temperature and $a_{\rm NR}$ is the scale factor at the time when all produced modes become non-relativistic. Here, for the sake of simplicity, we have also assumed an instantaneous transition from the relativistic regime to the non-relativistic one for the energy density of the entropy modes at the time when the Hubble expansion rate approaches the mass of the entropy modes $H_{\rm NR} = m_s$. Note that this transition should happen when the universe is radiation dominated, after the reheating and before the time of matter and radiation equality. 

During the radiation dominated era we have $H \propto \sqrt{g_{*,{\rm r}}} {a}^{-2}$. Neglecting changes in $g_{*,{\rm r}}$ for the times $H> m_s$, we find a simple expression $a_ {\rm e}/a_{\rm NR} \sim \mn_{\rm e}^{1/4}$, where we have used the definition of the parameter $\mn$ in Eq. \eqref{alpha-R} at the end of inflation. Although during an inflationary background the Hubble parameter and the conventional mass are considered to be constants with a good accuracy, we frequently use the notation $\beta_{\rm e}$ not only to be more precise but also to emphasize on the dependency of calculations to what happens at the end of inflation. 

Putting all things together, the relation \eqref{relic-DM} finally takes the following form
\begin{eqnarray}\label{relic-DM0}
\Omega_{\tiny s,0} =
{\cal O}(10^{20}) \mn_{\rm e}^{1/4}
\Big(\frac{T_{\rm r}}{10^{12}{\rm GeV}}\Big)\, 
\Omega_{\tiny s,{\rm e}} \,,
\end{eqnarray}
where we have substituted $T_0\sim10^{-13}$ GeV, $H_0\sim10^{-42}$ GeV, $g_{s*,0}=3.9$, and $g_{s*,{\rm r}}=106.75=g_{*,{\rm r}}$. 

Due to the curvature/entropy decomposition, by definition, there is no contribution from the entropy modes to the background at the beginning. However, the entropy modes can very efficiently be produced through the geometrical destabilization. This can destabilize the background trajectory of inflaton field as it is mentioned in \cite{Renaux-Petel:2017dia,Christodoulidis:2018qdw,Garcia-Saenz:2018ifx}. In our setup, this will happen when $\Omega_{\rm s,e} \sim {\cal O}(1)$. Thus, in order to have a consistent inflationary background, we have to make sure that the backreaction from the produced entropy modes will not spoil the background equations of motion (2.8), (2.9), and (2.15) which implies
\begin{align}\label{Constraint}
\Omega_{\tiny s,{\rm e}} \ll 1 \,.
\end{align}
In other words, destabilization of the background trajectory always comes with the overproduction of dark matter in our setup.
This is because we consider the case where the negative curvature of
the field space becomes more and more prominent towards the end of
inflation (see e.g. Eq. (5.1) below), unlike the original setup of
geometrical destabilization.

The large prefactor ${\cal O}(10^{20})$ in the result \eqref{relic-DM0} shows that if heavy modes with $\mn_{\rm e} \gtrsim {\cal O}(1)$ excite, then even small amount of particle production with $\Omega_{\tiny s,{\rm e}} \lesssim {\cal O}(10^{-20})$ is enough to obtain $\Omega_{\tiny s,0}={\cal O}(1)$. We will show that this is indeed possible in our model. On the other hand, for the light modes with $\mn_{\rm e}\lesssim {\cal O}(10^{-60})$ or equivalently $m_s\lesssim {\cal O}(10^{-8} {\rm eV})$, we need larger values of $\Omega_{\tiny s,{\rm e}} = {\cal O}(10^{-5})$ to achieve $\Omega_{\tiny s,0}={\cal O}(1)$.

The first criterion that our model should satisfy is to provide enough dark matter at the background level. This can be simply checked by looking at the accumulated energy density \eqref{rho-s}. However, in order to study the phenomenology of the produced dark matter particles, we also need to look at the scale dependence of the dark matter spectrum. Therefore, we define the dimensionless spectral density of the dark matter as
\begin{eqnarray}\label{Ps-def}
\Omega_{\tiny s} \equiv \int_{k_{\rm min}}^{k_{\rm max}} \dd \ln{k}\,\, P_s(k) \, . %\hspace{1.5cm}
%P_s(k) \equiv \frac{d\Omega_{\tiny s}}{d\ln{k}} \,.
\end{eqnarray}
The corresponding spectral tilt of entropy perturbations is given by
\begin{eqnarray}\label{n-s}
n_{\cal S} (k) - 1 \equiv \frac{\dd \ln{P_s}}{\dd\ln{k}} \,.
\end{eqnarray}
One can discriminate between different dark matter models by means of the spectral density $P_s(k)$ and the spectral tilt \eqref{n-s} as they characterize the dependence of the dark matter relic on the scale \cite{Graham:2015rva,Bastero-Gil:2021wsf}. Specifying an explicit functional form of the curvature of the field space ${\mathbb{R}}$, we can find explicit forms of $\Omega_{\tiny s}$, $P_s(k)$ and $n_{\cal S} (k)$ as we shall show in Sections \ref{sec-flat} and \ref{sec-curved}.

Moreover, we have to make sure that all excited entropy modes become non-relativistic before the time of matter and radiation equality. Thus, the conditions
\begin{equation}\label{NR-conditions}
m_s \gg H \,, 
\hspace{1cm} \mbox{and} \hspace{1cm} 
m_s \gg \frac{k_{\rm max}}{a} \,,
\end{equation}
should meet before the time of matter and radiation equality. From the above conditions, we find the following lower bound on the mass of the excited entropy modes \cite{Salehian:2020asa}
\begin{equation}\label{R-condition}
\mn_{\rm e} \gtrsim\max \Bigg\{
10^{-43} \, \Big(\frac{a_{\rm e} H_{\rm e}}{k_{\rm max}} \Big)^2 \,,
10^{-86} \, \Big( \frac{10^{12}\,{\mbox{GeV}}}{T_{\rm r}}\Big)^2
\Bigg\}
\times
\Big(\frac{10^{12}\,{\mbox{GeV}}}{T_{\rm r}}\Big)^2\,.
\end{equation} 
To obtain the above result, we have substituted $T_{\rm eq}\sim {\cal O}(\mbox{eV})$ for the matter-radiation equality temperature. 

Before closing this section, some comments are in order. First, we have considered the geodesic motion in the field space by assuming $\eta_{\bot}=0$ while the non-geodesic motion with $\eta_{\bot}\neq0$ is also an interesting possibility. For $\eta_{\bot}\neq0$, the negative contribution from the last term in the expression of $m_{\rm eff}^2$ in \eqref{effective-mass} would lead to the production of dark matter particles even for positive curvature ${\mathbb R}>0$. Secondly, although the curvature and entropy perturbations decouple at the linear level for $\eta_{\bot}=0$, they still interact with each other at the nonlinear level. At some point when the entropy modes are produced efficiently due to the geometrical destabilization, these nonlinear interactions are no longer negligible. This will open an indirect decay channel of the produced entropy modes to the SM particles. Therefore, assuming that this decay rate to be small, we find an upper bound on the mass of the entropy modes. This upper bound on the mass indicates that even if we excite superheavy entropy modes through the large negative values of the curvature of field space, they will decay to the SM particles. Thirdly, we have considered an instantaneous reheating for the sake of simplicity. Had we considered the standard reheating scenario during which $\dot{\sigma}$ oscillates, then so would $\alpha$ in Eq. \eqref{omega}. Consequently, the entropy modes would be excited even for positive curvature ${\mathbb R}>0$ through the process of parametric resonance. We leave the interesting cases of $\eta_{\bot}\neq0$ and the production of dark matter particles through parametric resonance during reheating phase for future studies.

\section{Flat field space}\label{sec-flat}

Let us first consider the simplest case of multiple field inflation with a flat field space. Considering ${\mathbb R}=0$ or equivalently $\mr=0$ in Eq. \eqref{omega}, the equation of motion for the entropy mode function takes the well-known form
\begin{align}
\label{omega-flat}
s_k'' + \omega_k^2\, s_k = 0 \;, \hspace{1.5cm}
\omega_k^2= k^2 
+ \frac{\mn}{\tau^2}
- \frac{2}{\tau^2} \,.
\end{align}
The positive frequency solution with the Bunch-Davies initial condition for the above equation is given by
\begin{equation}
\label{smod:ingoing}
s_k = \dfrac{\sqrt{-\pi \tau}}{2} \, {\rm H}_{\mu_1}^{(1)}(x)\,;
\hspace{1.5cm}  
\mu_1\equiv\sqrt{\frac{9}{4}-\mn} \,.
\end{equation}
There is no particle production when $\mn \gg1$ during inflation since the second term in $\omega_k^2$ in Eq. \eqref{omega-flat} is larger than the third term which implies $\omega_k^2>0$ all the time. On the other hand, for $\mn\ll1$ during inflation, we can have $\omega_k^2<0$ from the negative contributions due to the last term which is nothing but the gravitational particle production. In this case, we find $k_{\rm min}=-\sqrt{2}/\tau_{\rm i}$ and $k_{\rm max}=-\sqrt{2}/\tau_{\rm e}$, namely
\begin{equation}\label{limit-flat}
x_{\rm min} = - k_{\rm min} \tau_{\rm e} = \sqrt{2}\, e^{-{\cal N}} \,,
\hspace{1.5cm}
x_{\rm max} = - k_{\rm max} \tau_{\rm e} = \sqrt{2} \,,
\end{equation}
where we have used the fact that $\tau_{\rm i}/\tau_{\rm e}=e^{\cal N}$. Here, ${\cal N}$ is the total number of \textit{e}-folds during inflation. 

Substituting solution \eqref{smod:ingoing} into Eq. \eqref{Omega-s}, the spectral density defined in Eq. \eqref{Ps-def} turns out to be
\begin{eqnarray} \label{Ps-flat}
P_{s,{\rm e}}^{\rm flat}(x_{\rm e}) &=& \frac{H_{\rm e}^2}{48 \pi \Mp^2} \, x_{\rm e}^3
\Big[ x_{\rm e}^2 |H_{\mu_1+1}^{(1)}(x_{\rm e})|^2 +
\Big( \mn_{\rm e} + x_{\rm e}^2 + (\mu_1+3/2){}^2 \Big)
|H_{\mu_1}^{(1)}(x_{\rm e})|^2
\nonumber \\
&& \hspace{3.5cm}
- 2 x (\mu_1+3/2) {\rm Re} \big[ H_{\mu_1}^{(1)}(x_{\rm e}) H_{\mu_1+1}^{(2)}(x_{\rm e}) \big]
\Big] \,,
\end{eqnarray}
where $x_{\rm e} \equiv -k \tau_{\rm e}$. In the case of the flat field space, as it can be seen from Eq. \eqref{smod:ingoing}, there is only gravitational particle production due to the term $-2/\tau^2$ which only happens for the very light modes with $\mn\ll1$. Consequently, only superhorizon modes can be excited. This is similar to what happens for the curvature perturbations when they become classical after horizon crossing. 

Considering $\mn_{\rm e}\ll1$, we have $\mu_1\approx3/2$ and \eqref{Ps-flat} simplifies to
\begin{eqnarray}
P_{s,{\rm e}}^{\rm flat}(x_{\rm e}) \approx 
\frac{H_{\rm e}^2}{24 \pi^2 \Mp^2} x_{\rm e}^2 (1+2x_{\rm e}^2) \,.
\end{eqnarray} 
The above spectrum is blue-tilted with the peak $P_{s,{\rm e}}^{\rm flat}|_{x_{\rm e}=x_{\rm max}}=5H_{\rm e}^2/12\pi^2\Mp^2$ and the spectral tilt
\begin{eqnarray}\label{n-s-flat}
n_{{\cal S},{\rm e}}^{\rm flat}(x_{\rm e}) - 1 = 2 \bigg( \frac{1 + 4 x_{\rm e}^2}{1 + 2 x_{\rm e}^2} \bigg) \,.
\end{eqnarray}

For the corresponding accumulated energy density, from Eq. \eqref{Ps-def} we find
\begin{eqnarray}\label{C-flat-space}
\Omega_{\tiny s,{\rm e}}^{\rm flat} = 
\frac{H_{\rm e}^2}{24 \pi^2 \Mp^2} \int_{x_{\rm min}}^{x_{\rm max}} \dd x%_{\rm e} 
\, 
x%_{\rm e} 
\, \big( 1+2x%_{\rm e}
^2 \big) \, \approx \frac{H_{\rm e}^2}{8\pi^2 \Mp^2} \,,
\end{eqnarray}
where we have neglected contributions from the lower limit in comparison with the contributions from the upper limit.

Substituting the result \eqref{C-flat-space} into Eq. \eqref{relic-DM0}, we find the relic density of dark matter as follows
\begin{eqnarray}\label{relic-DM0-flat}
\Omega^{\rm flat}_{\tiny s,0}(\mn_{\rm e}) =
{\cal O}(10^{-6}) \mn_{\rm e}^{1/4}
\Big(\frac{T_{\rm r}}{10^{12}{\rm GeV}}\Big)^5
 \,,
\end{eqnarray}
where we have used the relation $3\Mp^2H_{\rm e}^2=(\pi^2/30)g_{*,{\rm r}}T_{\rm r}^4$ and have substituted $g_{*,{\rm r}}=106.75$. Note that although $\Omega_{\tiny s,{\rm e}}^{\rm flat}$ is very small, one can achieve $\Omega^{\rm flat}_{\tiny s,0}={\cal O}(1)$ for $T_{\rm r}\gtrsim{\cal O}(10^{13} {\rm GeV})$ for the mass range $10^{-16}\lesssim\mn_{\rm e}\lesssim10^{-4}$. 

The idea of obtaining dark matter from the entropy modes induced by the scalar isocurvature superhorizon modes is already discussed in the literature. In Ref. \cite{Polarski:1994rz}, it was pointed out that excited entropy modes can contribute to part of the observed dark matter relic density. More recently, it has been shown that considering a general spectator scalar field during inflation, it is possible to achieve the whole of dark matter \cite{Alonso-Alvarez:2018tus,Markkanen:2018gcw,Padilla:2019fju,Tenkanen:2019aij,Herring:2019hbe,Herring:2020cah,Cosme:2020nac,Ling:2021zlj}. In all of these models, superhorizon entropy modes are excited through the process of gravitational particle production. In the next subsection, we focus on the effects of the curvature of field space in particle production and find a new model of dark matter production in which subhorizon modes can be excited as well.

\section{Curved field space}\label{sec-curved}

In order to have an efficient particle production from the curvature of the field space, we need $\alpha-\beta\gtrsim{\cal O}(1)$ so $\omega_k^2 <0$ for these modes.  As we have shown in Eq. \eqref{alpha_bound}, the CMB bound on the isocurvature modes implies $\alpha-\beta\lesssim -0.1$ at the time when the CMB scales exit the horizon. Thus, it is not possible to have an efficient particle production via the curvature of the field space at the CMB scales. On the other hand, there is no bound at smaller scales and the curvature of the field space can be large so that $\alpha-\beta\gtrsim {\cal O}(1)$. Thus, we consider the case where the curvature of the field space ${\mathbb R}$ increases monotonically in time. This feature is quite general and we can consider many scenarios with different functional forms for the curvature of the field space which meet this criterion. In addition, we can also consider the situation where the mass $m_s$ changes during inflation such that $\alpha-\beta\gtrsim {\cal O}(1)$. This is possible, since the mass is related to $V_{NN}$ which may take complicated form in a curved field space. However, to simplify the analysis, we restrict our analysis to the situation where $m_s$ is constant during inflation while allowing for $\alpha$ to increase towards the final stage of inflation (i.e. after the time when the modes of CMB scales have left the horizon).

In order to have an explicit model which is analytically solvable, we consider the following simple power-law form
\begin{equation}\label{R-power-law}
{\mathbb R} = {\mathbb R}_{\rm e} \left(
\dfrac{\tau_{\rm e}}{\tau}\right)^{2p} \,;
\hspace{1.5cm}  p>0 \,,
\end{equation}
where ${\mathbb R}_{\rm e}<0$ is the curvature of field space at the end of inflation $\tau =\tau_{\rm e}$ and $p$ ($>0$) is a free parameter of the model. 
The specific value of $p$ is not important and in our analysis we may consider a typical value $p\gtrsim 1$.  The dimensionless parameter $\alpha$ defined in Eq. \eqref{alpha-R} is then a function of time given by
\begin{align}\label{alpha-model}
\mr = \mr_{\rm e} \left(
\dfrac{\tau_{\rm e}}{\tau}\right)^{2p} \,; \hspace{1.5cm}  
\mr_{\rm e} \equiv -\epsilon \Mp^2 \mathbb{R}_{\rm e}>0  \,.
\end{align}

The parameter $\mr_{\rm e}$ denotes the maximum value of $\mr$ at the end of inflation. Technically speaking the above ansatz  is applicable only after the time when modes of CMB scales have left the horizon  and until the end of inflation. This is because from the CMB constraint  on the entropy perturbations 
we require that $\alpha \lesssim -0.1 + \beta $ so for small $\beta$ the ansatz 
(\ref{alpha-model}) may not be consistent. Therefore in the following analysis 
we employ ansatz (\ref{alpha-model})  
after the time when the CMB scales have left the horizon till end of inflation. 
We further assume that $\mr\to0$ when the reheating is completed.
This may happen for example when  one field decays to radiation during reheating so we end up with a one dimensional field space which is flat by construction. 

Inserting Eq. \eqref{alpha-model} into Eq. \eqref{omega}, the equation of motion for the entropy mode function takes the following form
\begin{align}
\label{omega-model}
s_k'' + \omega_k^2\, s_k = 0 \;, \hspace{1.5cm}
\omega_k^2= k^2 
+ \frac{\mn}{\tau^2}
- \frac{2}{\tau^2}
-\dfrac{\mr_{\rm e}}{\tau^2}\left(\dfrac{\tau_{\rm e}}{\tau}\right)^{2p} \,.
\end{align}
To ensure that the effect of the curvature of the field space becomes important sometime during inflation, we consider
\begin{equation}
\mr_{\rm e} > |\mn-2| \,.
\end{equation}
In this case the last term in the right hand side of $\omega_k^2$ in Eq. \eqref{omega-model} dominates over the second and third terms towards the end of inflation. This happens at the time ${\tau}_{\rm c}$ given by
\begin{align}
\label{tau_c}
\dfrac{{\tau}_{\rm c}}{\tau_{\rm e}} = \left(
\dfrac{\mr_{\rm e}}{|\mn-2|}
\right)^{\frac{1}{2p}} \,.
\end{align}

Recasting the relation \eqref{tau_c} in term of the number of $e$-folds, we obtain
\begin{align}\label{deltaN}
\Delta {\cal N} \equiv \ln \left( \frac{{\tau}_{\rm c}}{\tau_{\rm e}}\right) = \dfrac{1}{2p} \ln \left(
\dfrac{\mr_{\rm e}}{|\mn-2|}
\right) \,,
\end{align}
where $\Delta {\cal N}$ denotes the \textit{e}-folding number for the period when the last term in \eqref{omega-model} dominates over the second and third terms before the end of inflation. The larger $\mr_{\rm e}$ is, the sooner the effect of the curvature of field space becomes important. 

As we already mentioned, we are interested in the modes with $\omega_k^2<0$ that can be excited by the negative curvature of the field space. In this case, from Eq. \eqref{omega-model} we find
\begin{equation}\label{tachyonic-curved}
\begin{cases}
x_{\rm min} = \sqrt{ (2-\mn_{\rm e}) \big( e^{-2p({\cal N}-\Delta{\cal N})} + 1 \big)} \, e^{-{\cal N}} \,,
&x_{\rm max} = \sqrt{ (2-\mn_{\rm e}) \big( e^{2p\Delta{\cal N}} + 1 \big) }  \,,
\hspace{2cm} \mn_{\rm e} < 2
\\
x_{\rm min} = 0 \,,
&x_{\rm max} = \sqrt{ (\mn_{\rm e}-2) \big( e^{2p\Delta{\cal N}} - 1 \big) }  \,,
\hspace{2cm} \mn_{\rm e} > 2
\end{cases}
\end{equation}
where we have used $\tau_{\rm c}/\tau_{\rm e}=e^{\Delta{\cal N}}$ and $\tau_{\rm i}/\tau_{\rm e}=e^{\cal N}$. 

Now, we only need to find solutions for the entropy mode function which satisfies Eq. \eqref{omega-model}. This equation cannot be solved analytically. However, based on the above discussion, we know that the last term in $\omega_k^2$ in Eq. \eqref{omega-model} can be neglected at early times $\tau \ll {\tau}_{\rm c}$ while it dominates towards the end of inflation $\tau \gg {\tau}_{\rm c}$. Solving separately for these two different phases and imposing the Bunch-Davies initial condition at $\tau \ll {\tau}_{\rm c}$, we find 
\begin{eqnarray}\label{s-curved}
s_k = \dfrac{\sqrt{-\pi \tau}}{2} \begin{cases}
 {\rm H}_{\mu_1}^{(1)}(x) \,,
& \hspace{1cm} \tau \ll {\tau}_{\rm c} \\
{\rm H}_{\mu_2}(iy) \,,
& \hspace{1cm} \tau \gg {\tau}_{\rm c}
\end{cases}
\end{eqnarray}
where ${\rm H}_{\mu_1}^{(1)}(x) $ is the Hankel function of the first kind and 
we have defined
\begin{eqnarray}\label{H-function}
{\rm H}_{\mu_2}(iy) \equiv 
{c}_2 \, {\rm H}_{\mu_2}^{(1)}(i y) + {c}_1 \, {\rm H}_{\mu_2}^{(2)}(i y) \,,
\end{eqnarray} 
with
\begin{align}\label{y}
y \equiv \dfrac{\sqrt{\mr_{\rm e}}}{p}\left(\dfrac{\tau_{\rm e}}{\tau} \right)^p \,, 
\hspace{1.5cm}
\mu_2 \equiv \frac{1}{2p} \,.
\end{align}
Matching the ingoing and outgoing solutions in \eqref{s-curved} at $\tau={\tau}_{\rm c}$, fixes the coefficients ${c}_{1,2}$. More specifically, demanding that both $s_k$ and $s_k'$ to be continuous at ${\tau}_{\rm c}$ yields\footnote{More precisely, we need to write the junction conditions for ${\cal F}_k$ and their conjugate momenta to avoid any discontinuity. However, we have ${\cal F}_k=a s_k$ and we assume that the scale factor and the Hubble expansion rate do not change significantly across the time $\tau=\tau_{\rm c}$.}
\begin{eqnarray}
\label{c12}
c_{\ell}(x_{\rm c}) = \frac{i \pi}{4 p} (-1)^{\ell} \left[  i p y_{\rm c} 
H_{\mu_1}^{(1)}(x_{\rm c}) \,
H_{\mu_2}^{(\ell)}{}'(iy_{\rm c}) + x_{\rm c} \,   H_{\mu_1}^{(1)}{}'(x_{\rm c}) \,  
H_{\mu_2}^{(\ell)}(iy_{\rm c}) \right] \, ,\hspace{1.5cm} \ell=1,2
\end{eqnarray}
where the primes denote derivatives with respect to the corresponding arguments. In the above results 
\begin{eqnarray}\label{yc}
{x}_{\rm c} = -k {\tau}_{\rm c} \,,
\hspace{1.5cm}
y_{\rm c}=\frac{\sqrt{\mr_{\rm e}}}{p} e^{-p\Delta{\cal N}} = 
\frac{\sqrt{|\mn-2|}}{p} \,.
\end{eqnarray}
It is worth noting that the coefficients $c_\ell$ depend on $k$ via the dependence on $x_{\rm c}$. 

\begin{figure}[t!]
	\begin{center}
		\includegraphics[scale=0.4]{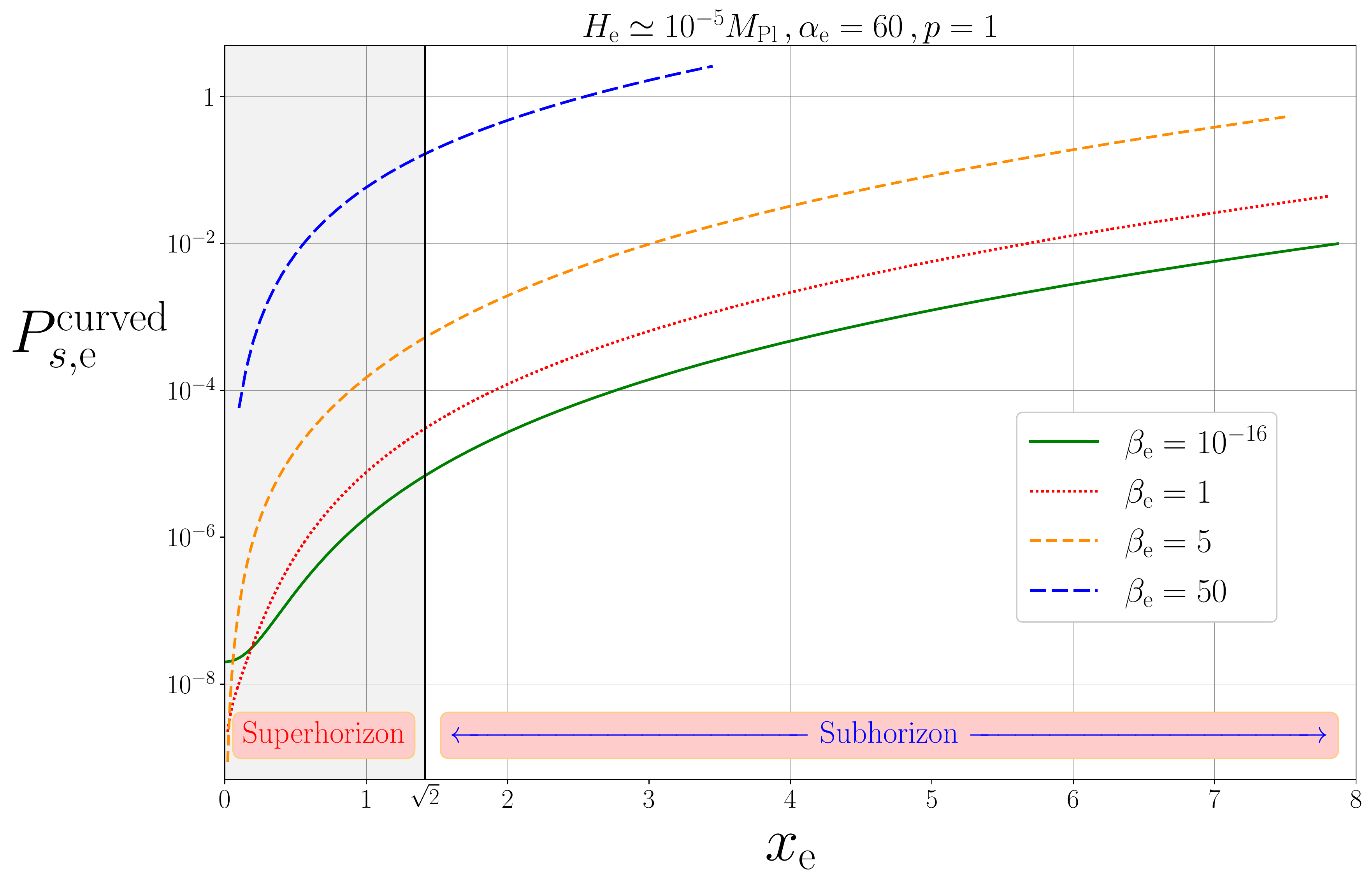}
		\caption{\footnotesize The spectral density versus the momentum is plotted for the excited entropy modes with different masses represented by parameter $\beta_{\rm e}$.  The superhorizon modes $x_{\rm e}<\sqrt{2}$ are enhanced by a higher rate than the subhorizon ones with $x_{\rm e}>\sqrt{2}$. However, the accumulated energy density $\Omega_{\tiny s,{\rm e}}^{\rm curved}$, given by Eq. \eqref{C-curved}, receives more contributions from the subhorizon modes.}
		\label{fig:Ps}
	\end{center}
\end{figure}

\begin{figure}[t!]
	\begin{center}
		\includegraphics[scale=0.4]{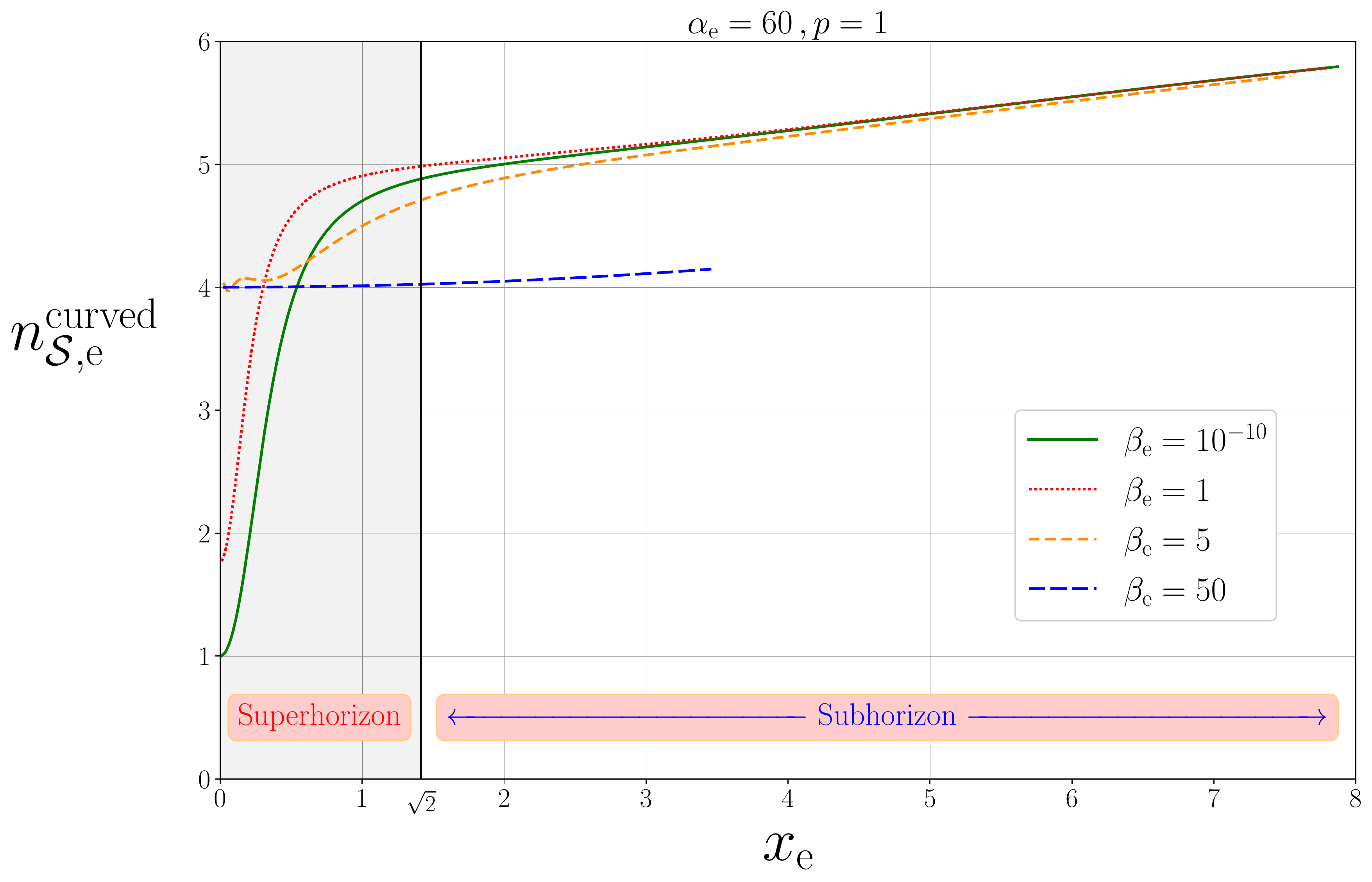}
		\caption{\footnotesize The tilt of the spectral density $n_{{\cal S}, {\rm e}}^{\rm curved}(x_{\rm e})-1 = \dd\ln{P^{\rm curved}_s}/\dd\ln{k}$ versus the momentum is plotted for the excited entropy modes with different masses. The spectrum is always blue-tilted while the rate of production for the heavy modes is less than the light ones (denoted by the solid green curve).}
		\label{fig:ns}
	\end{center}
\end{figure}

\begin{figure}[t!]
	\begin{center}
		\includegraphics[scale=0.4]{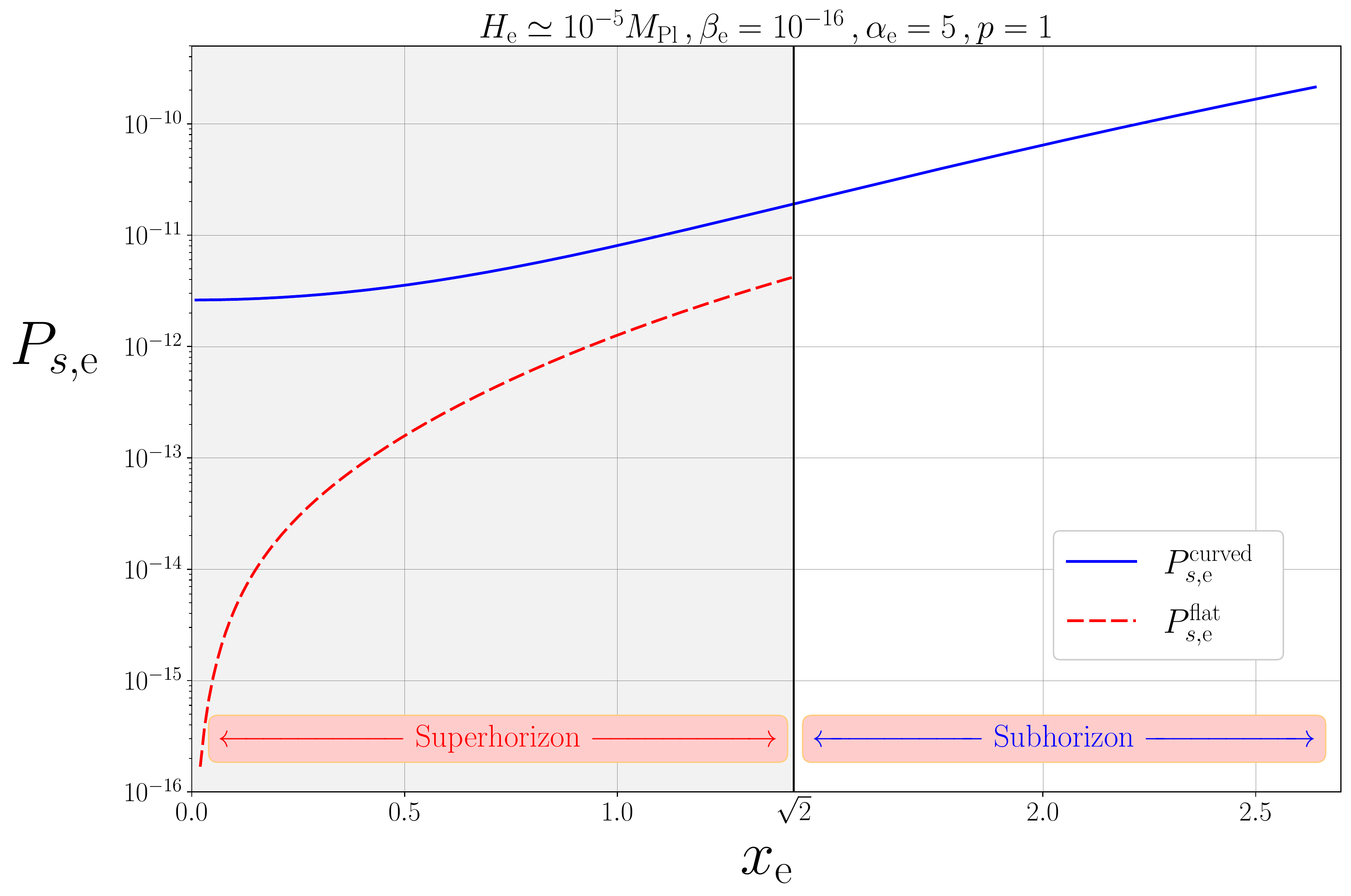}
		\caption{\footnotesize The spectral density for light entropy modes in the model with a flat field space versus the curved field space one. Only superhorizon modes $x_{\rm e}<\sqrt{2}$ can be excited in the case of flat field space while both superhorizon modes $x_{\rm e}<\sqrt{2}$ and subhorizon modes $x_{\rm e}>\sqrt{2}$ can be excited in the model with a curved field space. Thus, the location of the peaks for the spectral densities in the cases of flat and curved field spaces are different.}
		\label{fig:FvsC}
	\end{center}
\end{figure}

Having the mode function at the time $\tau\gg\tau_{\rm c}$ in hand, we can compute the spectral energy density \eqref{Ps-def} for entropy modes at the end of inflation
\begin{eqnarray}\label{Ps-curved}
P_{s,{\rm e}}^{\rm curved}(x_{\rm e}) = \frac{H_{\rm e}^2}{48\pi\Mp^2} x_{\rm e}^3 \bigg[
\Big{|} x_{\rm e} \frac{dH_{\mu_2}}{dx_{\rm e}} + \frac{3}{2} H_{\mu_2} \Big{|}^2 
+ \Big( x_{\rm e}^2 + \beta_{\rm e}+\alpha_{\rm e} \Big) | H_{\mu_2} |^2 
\bigg] \,.
\end{eqnarray}
The spectrum is defined for any mode while we are interested in the growing modes in the range defined in \eqref{tachyonic-curved}. As seen from Figs. \ref{fig:Ps} and \ref{fig:ns}, the spectral density \eqref{Ps-curved} is always blue-tilted and it is an increasing function for all modes $x_{\rm min} \leq x_{\rm e} \leq x_{\rm max}$ with arbitrary masses. In this regard, it always has a peak at $x_{\rm e}=x_{\rm max}$ which depends on the model parameters. Moreover, we have compared the results for the light excited entropy modes in flat and curved field spaces in Fig. \ref{fig:FvsC}. As seen from the figure, the main contribution to the spectral energy density in the flat field space \eqref{Ps-flat} also comes from $x_{\rm max}$ but $x_{\rm max}$ only includes superhorizon modes as shown in Eq. \eqref{limit-flat}. In the case of curved field space, on the other hand, $x_{\rm max}$ includes subhorizon modes as well thanks to the negative contribution of the curvature of field space to $\omega_k^2$. Therefore, the location of the peak of the density spectrum in our model is different than in models of  dark matter  with a flat field space which are based on the excitation of the superhorizon entropy modes. 

On the other hand, the location of peak for the vector dark matter models can be at small scales as the subhorizon modes can be excited in some of these models as well \cite{Bastero-Gil:2018uel,Nakayama:2019rhg,Nakayama:2020rka,Nakai:2020cfw,Salehian:2020asa,Firouzjahi:2020whk,Bastero-Gil:2021wsf}. A question then may arise whether our model is distinguishable from the vector dark matter models or not. First, the shape of the spectral energy density in our model can be different than those in the vector dark matter models. This criterion is, however, model-dependent and one may expect a vector dark matter model which provides the spectral density with a shape more or less similar to our model. However, there is a fundamentally different criterion which makes the models based on the scalar isocurvature modes like our model, different than vector dark matter models. All vector dark matter models will provide vector perturbations which lead to small scale anisotropies while this is not the case for the scalar dark matter models \cite{Cembranos:2016ugq,Nomura:2019cvc,Tsukada:2020lgt,Miravet:2020kuj}. 

Therefore, based on the above mentioned points, our model is distinguishable not only from the scalar dark matter models based on the flat field space but also from the vector dark matter scenarios if one looks at both the location of the peak of the spectral density and small scale anisotropies.

Having the spectral density \eqref{Ps-curved} in hand, we can find the corresponding relic density in the model with a curved field space through the definitions \eqref{relic-DM0} and \eqref{Ps-def} as follows
\begin{eqnarray}\label{relic}
\Omega_{\tiny s,0}^{\rm curved}(\mr_{\rm e},p,\mn_{\rm e}) =
{\cal O}(10^{20})
\mn_{\rm e}^{1/4} \Big(\frac{T_{\rm r}}{10^{12}{\rm GeV}}\Big)\, 
\Omega_{\tiny s,{\rm e}}^{\rm curved}(\mr_{\rm e},p,\mn_{\rm e}) \,,
\end{eqnarray} 
where
\begin{eqnarray}\label{C-curved}
\Omega_{\tiny s,{\rm e}}^{\rm curved}(\mr_{\rm e},p,\mn_{\rm e}) 
= \int_{x_{\rm min}}^{x_{\rm max}} \dd \ln{x%_{\rm e}
}\,  \, P_{s,{\rm e}}^{\rm curved}(x%_{\rm e}
)
\,.
\end{eqnarray} 

\begin{figure}[t!]
	\begin{center}
		\includegraphics[scale=0.4]{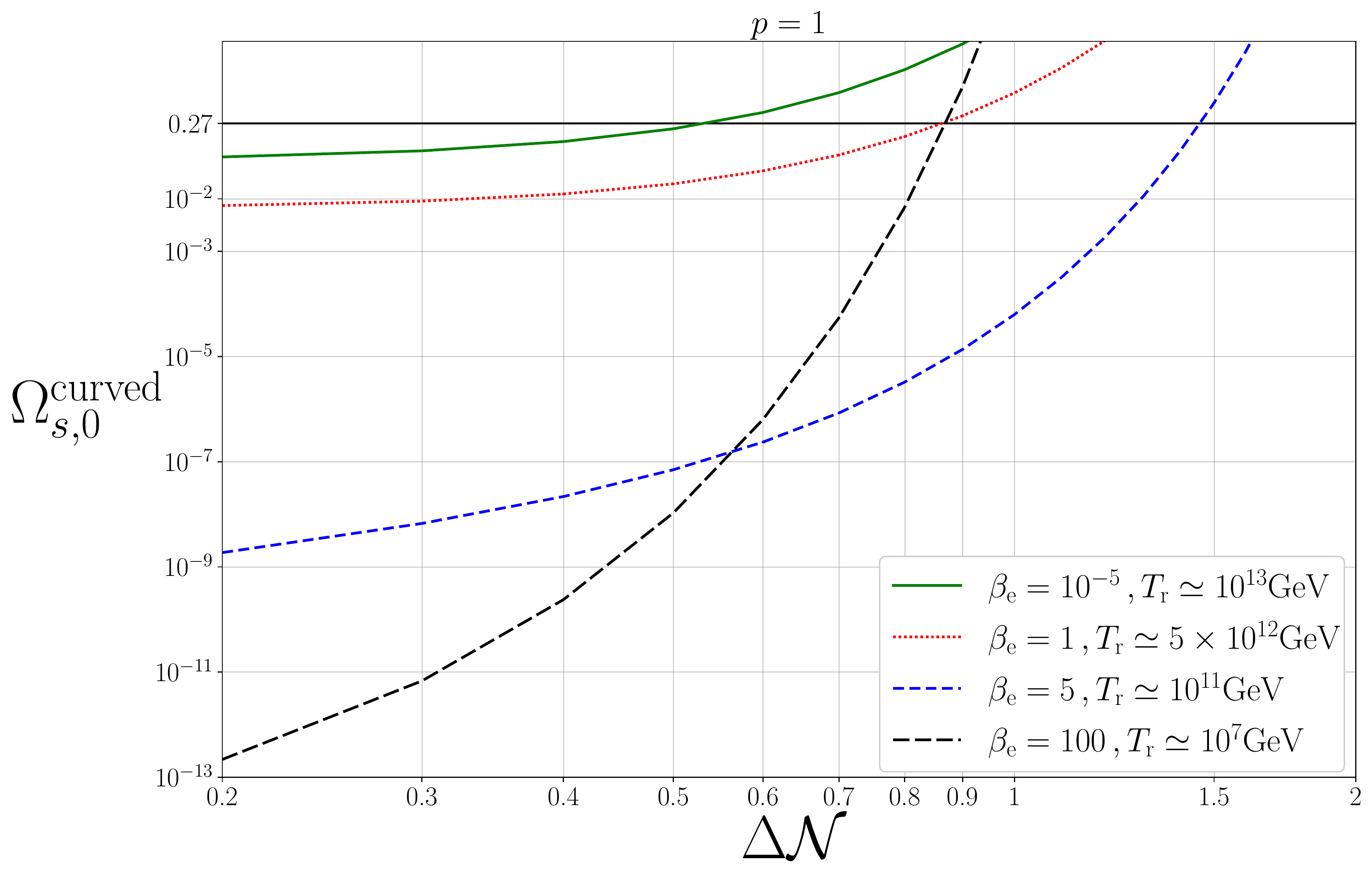}
		\caption{\footnotesize The accumulated energy density of the excited entropy modes versus the number of \textit{e}-folds $\Delta{\cal N}$, defined in \eqref{deltaN}, is plotted for different masses and different inflationary energy scales. The solid horizontal line shows the desired value of the dark matter relic density today, $\Omega_0=0.27$. 
The particle production is very efficient as $\Omega_{\tiny s,0}^{\rm curved} \sim {\cal O}(1)$ can be achieved even for $\Delta {\cal N} \lesssim {\cal O}(1)$.}
		\label{fig:Omega-s}
	\end{center}
\end{figure}

\begin{figure}[t!]
	\begin{center}
		\includegraphics[scale=0.4]{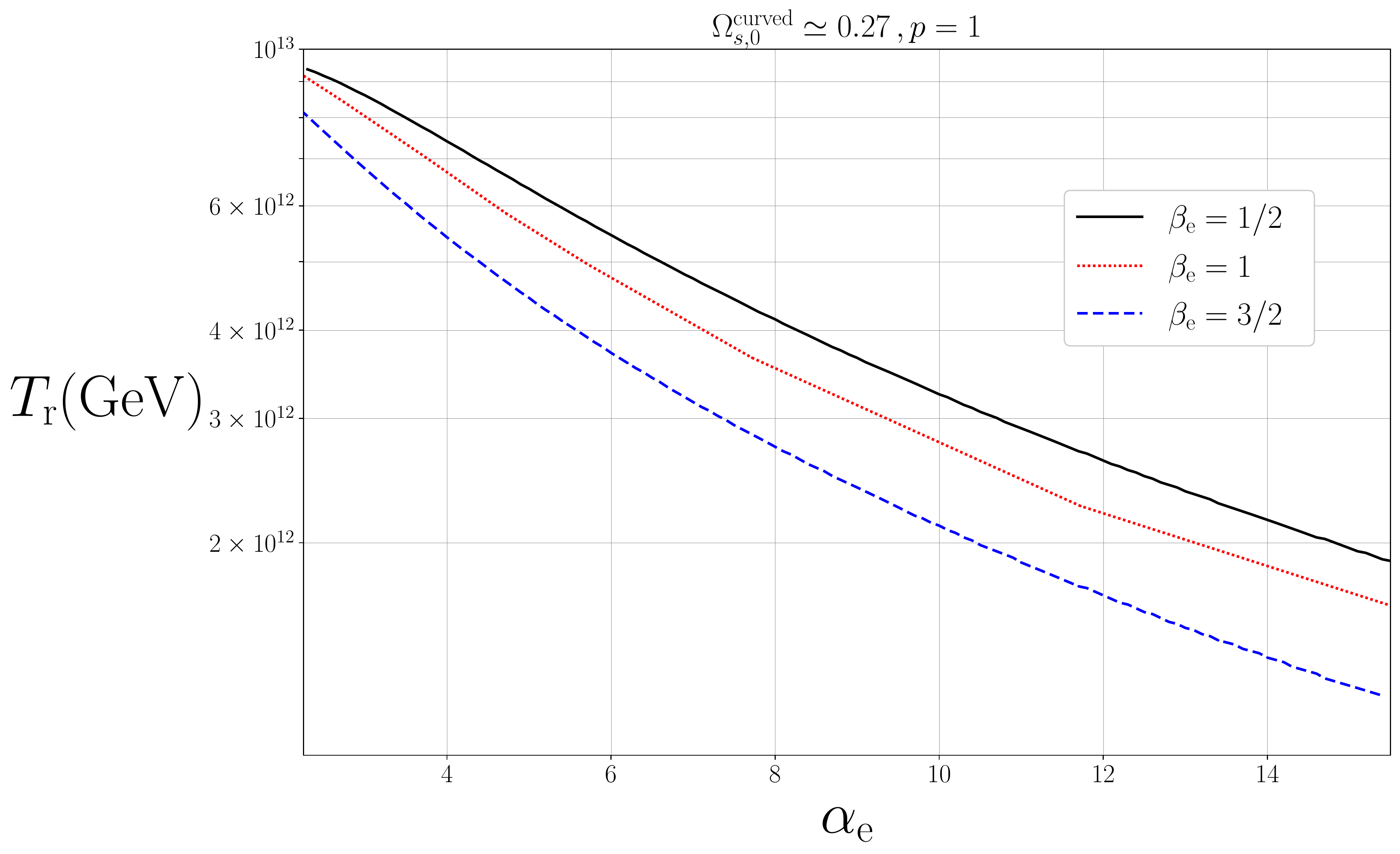}
		\caption{\footnotesize The curves show the suitable parameter space for $T_{\rm r}-\alpha_{\rm e}$ demanding that the entropy modes with semi-heavy masses $\beta_{\rm }=1/2,~1,~3/2$ provide all the dark matter in the universe.}
		\label{fig:Tr-alpha}
	\end{center}
\end{figure}

\begin{figure}[t!]
	\begin{center}
		\includegraphics[scale=0.4]{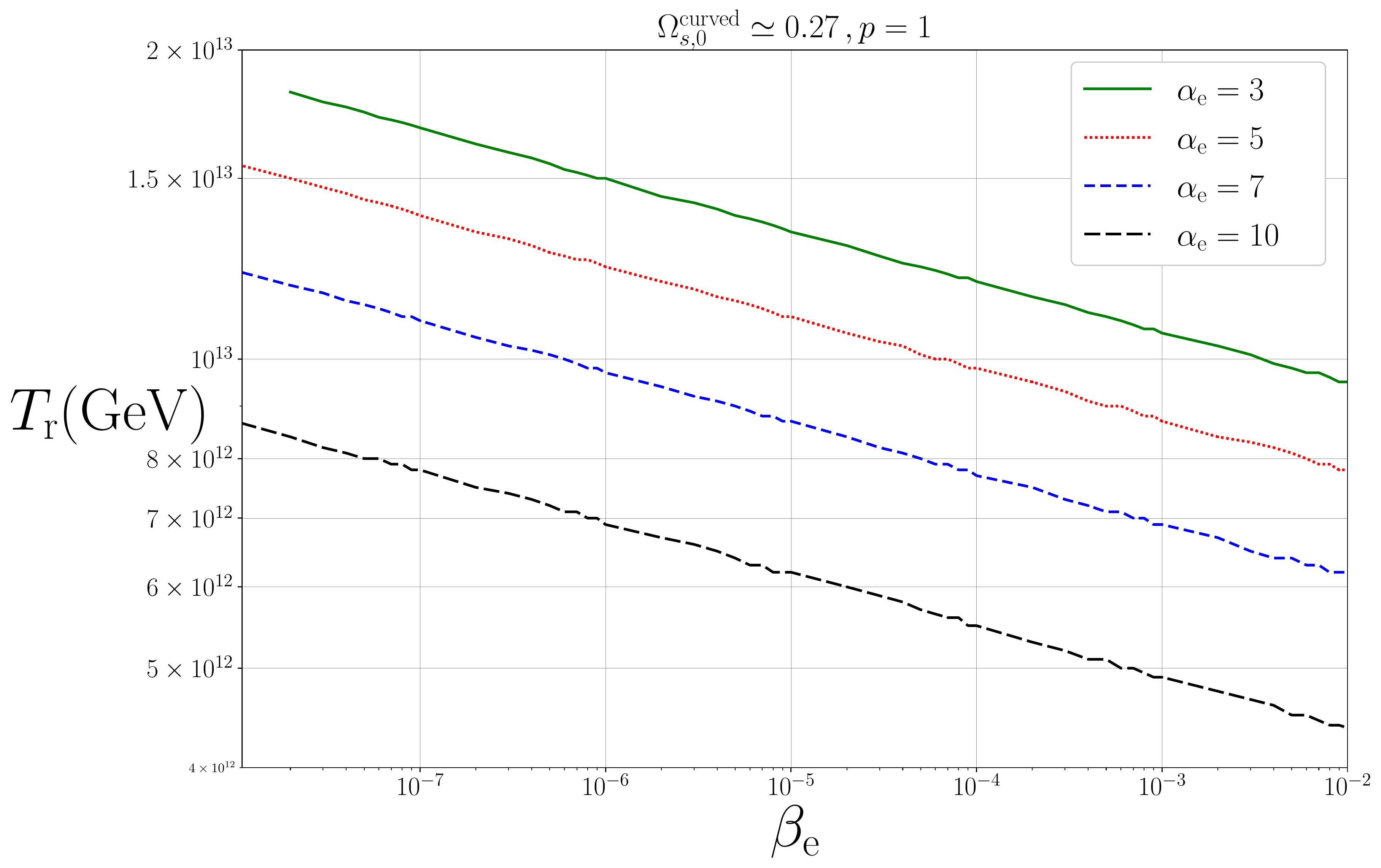}
		\caption{\footnotesize The curves show the suitable parameter space for $T_{\rm r}-\beta_{\rm e}$ with $\alpha_{\rm e}=3,~ 5,~ 7,~10$, demanding that the entropy modes with small masses generate all the dark matter in the universe.}
		\label{fig:Tr-beta}
	\end{center}
\end{figure}

We have presented the results for the relic \eqref{relic} in Fig. \ref{fig:Omega-s} for entropy modes with different masses in terms of $\Delta {\cal N}$  with  $p=1$. Demanding that the relic \eqref{relic} contains the total observed dark matter density, i.e. $\Omega_{\tiny s,0}^{\rm curved} \simeq 0.27$, leads to a parameter space for $\alpha_{\rm e}, \beta_{\rm e}$ and $T_{\rm r}$. We have presented in Figs.~\ref{fig:Tr-alpha} and~\ref{fig:Tr-beta} the suitable parameter space for $T_{\rm r}$, which represents the energy scale of inflation, in terms of $\alpha_{\rm e}$ and $\beta_{\rm e}$, respectively.

In the case of light ($\beta_{\rm e} \ll 1$) and semiheavy ($\beta_{\rm e} \sim 1$) entropy modes, we can achieve the right amount of dark matter for a wide range of the parameter space. On the other hand, for the heavy modes  ($\beta_{\rm e} \gg 1$)
the value of the parameters $\mn_{\rm e}$ and $\mr_{\rm e}\approx \mn_{\rm e} e^{2p\Delta{\cal N}}$ should be chosen very close to each other in order to produce the right 
amount of dark matter.  This shows that we need a fine-tuning for the heavy modes. Indeed, one can directly confirm that the larger $\mn_{\rm e}$ is, the more accurate the required fine-tuning is in order to achieve the right amount of dark matter  (i.e. not to overproduce dark matter).  This fact can be understood as follows. 

First, having large masses, from \eqref{relic} we see that we do not need to produce so many heavy dark matter particles and a small spectral density can do the job.  Secondly, but more importantly, we note that for the heavy entropy modes $\beta\gg1$, we have to assume $\alpha\gg1$ in order to be able to achieve the desired tachyonic condition $\omega_k^2<0$. Thus, we can neglect the term $-2/\tau^2$ in Eq. \eqref{omega} for $\beta\gg1$ and $\alpha\gg1$. The mass term $+\mn/\tau^2$ dominates at $\tau \ll \tau_{\rm c}$ while the term $-\mr/\tau^2$ dominates at $\tau \gg \tau_{\rm c}$. If we consider $\mn>\mr$, there would be no particle production. On the other hand, if we consider $\mn<\mr$, there would be a huge negative contribution to $\omega_k^2$ for some interval in the regime $\tau > \tau_{\rm c}$ which leads to the overproduction of dark matter. Based on the above two mentioned points, the only way to achieve the right amount of dark matter is to fine-tune the values of the mass term and the curvature of  field space such that $\mr \gtrsim \mn$ to produce a little amount of heavy dark matter particles in a very short time interval. In this regard, although it is in principle possible to produce the right amount of dark matter from the growing heavy entropy modes but we need to fine-tune the values of the parameters.

%%%%%%%%%%%%%%%%%%%%%%%
%%%%%%%%%%%%%%%%%%%%%%%

We have performed the integration in \eqref{C-curved} numerically as it is not possible to find an analytical result in general. In order to better understand the behaviour of the dark matter relic density \eqref{relic}, we now study the cases of heavy, semiheavy and light modes separately. Then, we will be able to find analytical expressions for the corresponding dark matter relic density.

\subsection{Heavy modes}

For the heavy entropy modes with $\mn \gg 1$ and $\alpha=\mathcal{O}(1)$, the term $+\mn/\tau^2$ dominates over all the other terms in Eq. \eqref{omega} during the whole period of inflation and, obviously, we do not have any particle production since  always $\omega_k^2>0$. However, if we consider $\alpha \gg1$ in addition to $\mn \gg1$, in principle, it is possible to satisfy the desired condition $\omega_k^2<0$. In this case, one can completely neglect the term $-2/\tau^2$  in \eqref{omega}, which is responsible for the gravitational particle production, in comparison with $+\mn/\tau^2$ and $-\mr/\tau^2$ terms. Thus, we end up with the conclusion that even heavy entropy modes can be excited through the geometrical destabilization mechanism \cite{Renaux-Petel:2015mga}. Note that this is not possible in the case of a flat field space ($\alpha =0$) as $\mn \gg1$ implies $\omega_k^2> 0$ in Eq. \eqref{omega-flat}
for the whole period of inflation. 

Since $\alpha$ in our model \eqref{alpha-model} increases with time, the mass term $\mn/\tau^2$ dominates up to the time $\tau_{\rm c}$ defined in \eqref{tau_c}. Thus, there is no particle production for $\tau<\tau_{\rm c}$. For $\tau>\tau_{\rm c}$, the last term in Eq. \eqref{omega-model} dominates which gives negative contribution and makes the tachyonic growth with $\omega_k^2<0$ possible. The range of growing modes in \eqref{tachyonic-curved} for the heavy modes are characterized by
\begin{equation}\label{tachyonic-heavy}
x_{\rm min} = 0 \,,
\hspace{1.5cm}
x_{\rm max} \approx \sqrt{\mn_{\rm e}} \sqrt{e^{2p\Delta{\cal N}} - 1} \,.
\end{equation}
For the heavy entropy modes with $\mn_{\rm e}\gg1$, from Eqs. \eqref{smod:ingoing} and \eqref{yc} we see that $y_{\rm c}\gg1$ and $\mu_1=i\sqrt{\mn_{\rm e}}$. Therefore expanding \eqref{c12} for $y_{\rm c}\gg1$ and then using $H_{\nu}^{(1)}(z) \sim -i \sqrt{\frac{2}{\pi\nu}} \Big(\frac{{\rm e} z}{2 \nu}\Big)^{-\nu}$ for $\nu\gg1$, we find
\begin{equation}\label{c12-R-heavy}
c_{\ell} \approx 
\frac{ -i}{\sqrt{2p}}
e^{ \frac{(-1)^\ell }{p}\sqrt{\beta }}
e^{-\frac{\pi\sqrt{\beta } }{2}}
\left( \frac{e x_{\rm c}}{2\sqrt{\beta}} \right)^{-i \sqrt{\beta }}
\,.
\end{equation}
From the above result we see that $c_1/c_2 = e^{-(2/p)\sqrt{\beta}}\ll1$ for $\beta\gg1$. Thus, we ignore the contribution from $c_1$ in \eqref{H-function}. Substituting \eqref{c12-R-heavy} in \eqref{H-function} and using the result in \eqref{Ps-curved}, from Eqs. \eqref{relic} and \eqref{C-curved} we obtain 
\begin{eqnarray}\label{C-heavy}
{\Omega}_{\tiny s,{\rm e}}^{\rm H}(\Delta{\cal N},p,\mn_{\rm e}) \approx 
\frac{13 H_{\rm e}^2}{720\pi^2\Mp^2} 
e^{p\Delta{\cal N}} \big( e^{2p\Delta{\cal N}} - 1 \big)^{3/2}
e^{-\kappa \sqrt{\mn_{\rm e}}} 
 \mn_{\rm e}^2 
\,; \hspace{1.3cm}
\kappa \equiv \frac{2}{p}\big(e^{p\Delta{\cal N}}+\frac{\pi{p}}{2}-1\big) \,,
\end{eqnarray}
where we have used the relation \eqref{deltaN}. 

\begin{figure}[t!]
	\begin{center}
		\includegraphics[scale=0.4]{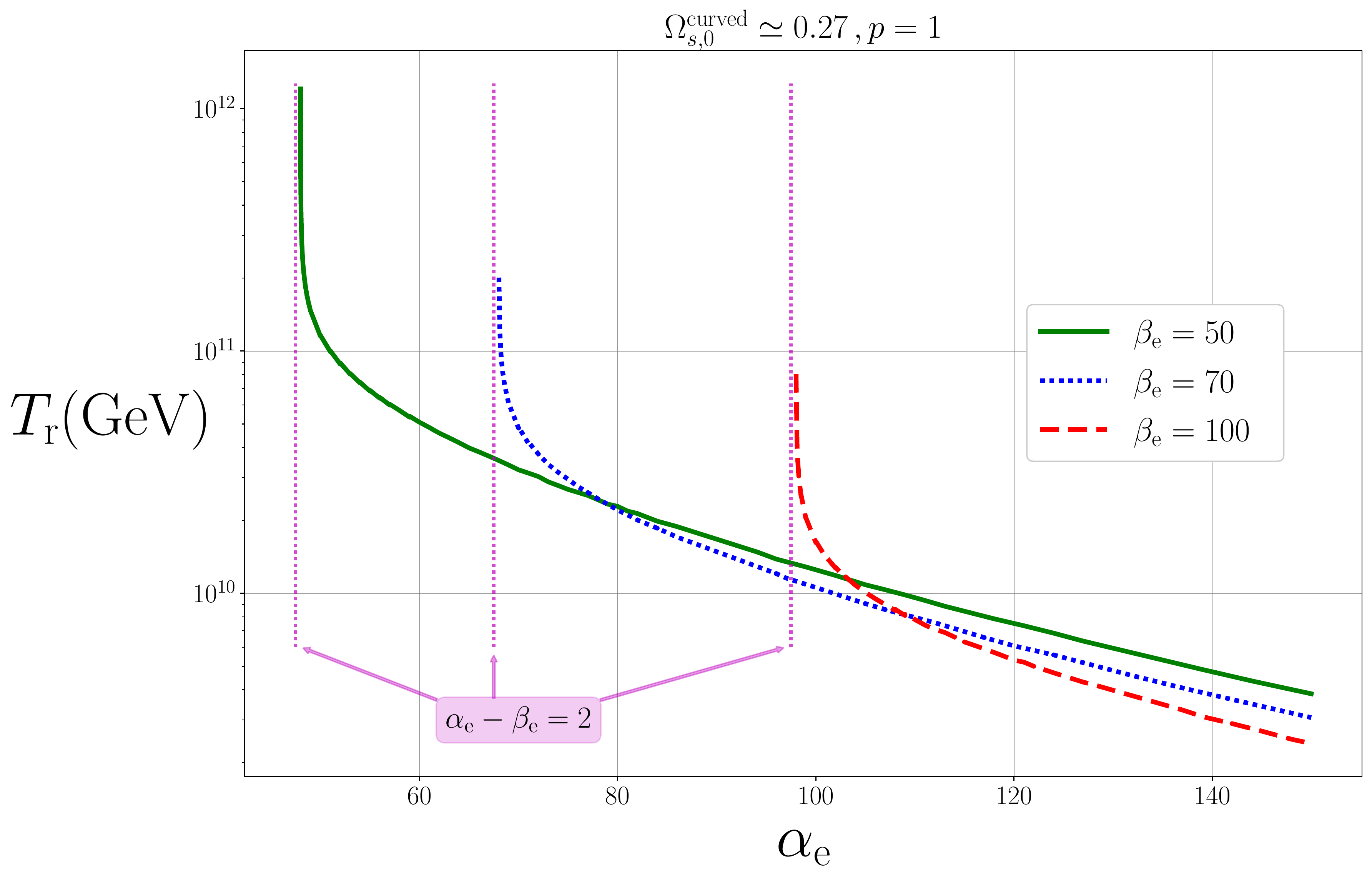}
		\caption{\footnotesize The curves show the suitable parameter space for $T_{\rm r}-\alpha_{\rm e}$, demanding that the entropy modes with heavy masses $\beta_{\rm e}=50,~70,~100$ generate all the dark matter in the universe. The requirement of high energy inflation, $T_{\rm r} > {\cal O}(10^{11} {\rm GeV})$, leads to the 
		fine-tuning $\alpha_{\rm e} \simeq \beta_{\rm e} -2$ indicated by the dotted vertical lines.}
		\label{fig:Tr-alpha-Heavy}
	\end{center}
\end{figure}

The corresponding energy density today is given by
\begin{eqnarray}\label{relic-heavy}
\Omega_{\tiny s,0}^{\rm H}(\Delta{\cal N},p,\mn_{\rm e}) =
{\cal O}(10^{20})
\mn_{\rm e}^{1/4} \Big(\frac{T_{\rm r}}{10^{12}{\rm GeV}}\Big)\, 
\Omega_{\tiny s,{\rm e}}^{\rm H}(\Delta{\cal N},p,\mn_{\rm e}) \,.
\end{eqnarray} 
As we have $\kappa>0$ for the typical values of $p$ and $\Delta{\cal N}$, the factor $e^{-\kappa \sqrt\mn_{\rm e}} = e^{-\kappa m_s/H_{\rm e}}$ in \eqref{C-heavy} is responsible for the well-known Boltzmann suppression factor for the heavy modes.  On the other hand, this mass suppression can be compensated by the factor $ \mn_{\rm e}^2$ which is originated from the curvature of the field space. This is the analytical confirmation of the existence of a fine-tuning that we mentioned at the end of the previous subsection. However, the fine-tuning issue for this model can be somehow ameliorated at the cost of lowering the energy scale of inflation as can be seen  in Fig.~\ref{fig:Tr-alpha-Heavy}.

%%%%%%%%%%%%%%
%%%%%%%%%%%%%%

\subsection{Semiheavy modes}

Now, we turn to the most interesting case of semiheavy entropy modes with $\mn_{\rm }={\cal O}(1)$. The critical value $y_{\rm c}$ defined in Eq. \eqref{yc} satisfies $y_{\rm c}\lesssim{1}$ for the semiheavy and light entropy modes with $\mn\lesssim{\cal O}(1)$ and the typical value of $p={\cal O}(1)$. However, we have numerically confirmed that for $0.5 \lesssim {y}_{\rm c} \lesssim {1}$, we can obtain approximate results by taking the limit $y_{\rm c}\ll1$ for the semiheavy entropy modes with very good accuracy. In this case, the coefficient $c_{1,2}$ in \eqref{c12} can be approximated as follows
\begin{equation}\label{c12-approx}
c_{1,2}(x_{\rm c}) \approx \frac{i^{-\mu_2}}{4} \mu_2 \Gamma(\mu_2)
\left( (2 \mu_1 + 1 ) H_{\mu_1}^{(1)}(x_{\rm c}) 
- 2 x_{\rm c} H_{\mu_1-1}^{(1)} (x_{\rm c})\right)
\bigg( \frac{2}{y_{\rm c}} \bigg)^{\mu_2} \,, \hspace{1.5cm} \mbox{for} \hspace{.5cm}
y_{\rm c} \ll 1 \,.
\end{equation}

As $c_1\simeq c_2$ for $y_{\rm c} \ll 1$, the solution \eqref{s-curved} at $\tau\gg\tau_{\rm c}$ simplifies to
\begin{align}
\label{smod}
s_k = e^{i\theta} \, \sqrt{-\pi \tau} \, c_1 \,J_{\mu_2} (i y) \,, 
\hspace{1.5cm} \tau \gg {\tau}_{\rm c}\,,
\end{align}
where $J_{\mu}(x)$ denotes the Bessel function and $\theta$ is a phase factor. Substituting the above result in Eq.~\eqref{rho-s} we find
\begin{eqnarray}\label{C-general0}
\Omega_{\tiny s,{\rm e}}^{\rm SH} 
= \frac{H_{\rm e}^2}{12 \pi^2 \Mp^2} \int_{x_{\rm min}}^{x_{\rm max}} \dd x%_{{\rm e}} 
\, 
\Big( L_{\rm e}
+ \left(\mn_{\rm e} + \mr_{\rm e} +x^2 \right) K_{\rm e} \Big) 
x^2 \, |c_1(x%_{\rm e}
)|^2 \,,
\end{eqnarray}
where we have defined $L(\tau)$ and $K(\tau)$ which are functions of time and not momentum as follows
\begin{equation}
L(\tau) \equiv \pi \big[
\sqrt{\alpha_{\rm e}} I_{1+\mu_2}(y) - I_{\mu_2}(y)
\big]^2 \,, \hspace{1.5cm}
K(\tau) \equiv \pi I_{\mu_2}(y)^2 \,.
\end{equation}
Here, $I_{\mu}(x) = i^{-\mu} J_{\mu}(x)$ is the modified Bessel function of the first kind. The limits of the integration in \eqref{C-general0} are given by \eqref{tachyonic-curved}.

Substituting Eq.~\eqref{c12-approx} in Eq.~\eqref{C-general0} and performing the integral, we find the following analytical result
\begin{align}\label{C-general}
\Omega_{\tiny s,{\rm e}}^{\rm SH} = \frac{H_{\rm e}^2}{3 \Mp^2}
\left(\dfrac{\Gamma(1+\frac{1}{2p})}{4{\pi}e^{\Delta{\cal N}}}\right)^2\left(\dfrac{2p}{\sqrt{\alpha_{\rm e}}}\right)^{1/p}
\bigg[\ 
\big( L_{\rm e}+\left(\mr_{{\rm e}}+\mn_{\rm e} \right)K_{\rm e} \big){\cal F}^{(2)}(\mu_1) 
+ e^{-2\Delta{\cal N}} K_{\rm e} \  {\cal F}^{(4)}(\mu_1)
\bigg] \,,
\end{align}
where
\begin{align}
{\cal F}^{(n)}(\mu_1) &\equiv \left(\mu_1+\dfrac{1}{2}\right)^2 {\cal I}^{(n)}_{\mu_1,\mu_1}+ {\cal I}^{(n+2)}_{\mu_1-1,\mu_1-1} 
- 2 \left(\mu_1+\dfrac{1}{2}\right) \ {\rm Re}\Big[{\cal I}^{(n+1)}_{\mu_1,\mu_1-1}\Big] \,;
\\
{\cal I}^{(n)}_{a,b} & \equiv 
\int_{x_{\rm min} e^{\Delta{\cal N}}}^{x_{\rm max} e^{\Delta{\cal N}}} \dd x%_{\rm c} 
\,
x%_{\rm c}
^n \ H^{(1)}_a(x%_{\rm c}
) \ H^{(2)}_b(x%_{\rm c}
) \,.
\end{align}
The integrals for ${\cal I}^{(n)}_{a,b}$ can be explicitly computed in terms of the Hypergeometric and Gamma functions.

Substituting Eq. \eqref{C-general} in Eq.~\eqref{relic-DM0}, we find the following result for the relic density of the dark matter produced by the semiheavy entropy modes
\begin{eqnarray}\label{relic-Sheavy}
\Omega_{\tiny s,0}^{\rm SH}(\mr_{\rm e},p,\mn_{\rm e}) =
{\cal O}(10^{20})
\mn_{\rm e}^{1/4} \Big(\frac{T_{\rm r}}{10^{12}{\rm GeV}}\Big)\, 
\Omega_{\tiny s,{\rm e}}^{\rm SH}(\mr_{\rm e},p,\mn_{\rm e}) \,.
\end{eqnarray} 
To simplify the result, we note that $x_{\rm min}\ll1$. Expanding \eqref{C-general} for the small values of $x_{\rm e}$, we find that the leading term is given by $x_{\rm e}^{3-2\mu_1}$. Therefore for the case of real $\mu_1$ with $\mn<9/4$, which includes semiheavy modes with $\mn={\cal O}(1)<9/4$, we always have $\mu_1<3/2$ and, therefore, contributions from the lower limit $x_{\rm min}$ are negligible. The dominant contributions are then given by the upper limit $x_{\rm max}$. From Eq. \eqref{tachyonic-curved} we find that $x_{\rm max}\gtrsim{\cal O}(1)$. For $x_{\rm max}\sim{\cal O}(1)$ we cannot simplify the result further  while we find the following simple result for $x_{\rm max}\gg{\cal O}(1)$
\begin{eqnarray}\label{C-Sheavy}
\Omega_{\tiny s,{\rm e}}^{\rm SH}(\mr_{\rm e},p,\mn_{\rm e}) \approx \frac{H_{\rm e}^2}{3 \Mp^2}
\frac{ \Gamma \left(1/2 p\right)^2 }{ {2}^7 \pi^3 p^2}
\left(\frac{2p}{\sqrt{\mr_{\rm e}}}\right)^{1/p} 
K_{\rm e} \, e^{\Delta{\cal N}}
x_{\rm max}^5\, {\cal T}(x_{\rm max};\mu_1) \,,
\end{eqnarray}
where we have defined
\begin{eqnarray}
{\cal T}(x;\mu_1) \equiv 
\frac{7-4 \mu_1}{5} \cot \left(\pi  \mu _1\right) 
\cos \left(2 x e^{\Delta \mathcal{N}} \right)- \sin \left(2 x e^{\Delta \mathcal{N}} \right) \,.
\end{eqnarray}

We can further simplify the result \eqref{C-Sheavy} by taking the limit $\lim_{x\to\infty}{\cal T}(x;\mu_1)$ which gives
\begin{eqnarray}\label{T-bound}
{\cal T}(x_{\rm max};\mu_1) \leq 
\bigg{|}\frac{5+(7-4 \mu _1) \cot \left(\pi  \mu_1\right)}{5\sin \left(\pi  \mu _1\right)} \bigg{|} \,.
\end{eqnarray}
In particular, we have ${\cal T}(x_{\rm max};\mu_1) \leq 1$ for $\mu_1=1/2$ or $\mn_{\rm e}=2$. The above result is valid for the whole range $0\leq{\mu_1}\leq\sqrt{7}/4$ or equivalently $1/2\leq{\mn_{\rm e}}\leq9/4$ while the right hand side of Eq. \eqref{T-bound} can be very large for small values of $\mu_1$. Substituting the above result in Eq. \eqref{C-Sheavy}, we find the following upper bound
\begin{eqnarray}\label{C-Sheavy-S}
\Omega_{\tiny s,{\rm e}}^{\rm SH}(\mr_{\rm e},p,\mn_{\rm e}) \lesssim \frac{H_{\rm e}^2}{3 \Mp^2}
\frac{ \Gamma \left(1/2 p\right)^2 }{ {2}^7 \pi^3 p^2}
\left(\frac{2p}{\sqrt{\mr_{\rm e}}}\right)^{1/p} 
K_{\rm e} \, e^{\Delta{\cal N}}
(2-\mn_{\rm e}+\mr_{\rm e})^{5/2}\,
\bigg{|}\frac{5+(7-4 \mu _1) \cot \left(\pi  \mu_1\right)}{5\sin \left(\pi  \mu _1\right)} \bigg{|} .
\end{eqnarray}
The main contributions come from the upper limit of the integral and we have shown that the energy density spectrum is blue-tilted. This is the reason why we have treated the result \eqref{C-Sheavy-S} for $x_{\rm max}\gg{\cal O}(1)$ to be an upper bound for all semiheavy modes. The result \eqref{C-Sheavy-S} was obtained for $1/2\leq{\mn}_{\rm }\leq 9/4$ and it is not valid for either $\mn_{\rm }\ll1$ or $\mn_{\rm }\gg1$.

\subsection{Light modes}

In the case of light dark matter $\mn\ll1$, we can neglect $\mn$ in Eq. \eqref{omega-model} and, therefore, we have particle production during the whole period of inflation $\tau_{\rm i}<\tau<\tau_{\rm e}$. The limits in \eqref{tachyonic-curved} simplify to
\begin{equation}\label{tachyonic-light}
x_{\rm min} \approx \sqrt{2} \, e^{-{\cal N}} \,,
\hspace{1.5cm}
x_{\rm max} = \sqrt{2} \sqrt{ e^{2p\Delta{\cal N}} + 1 } \,.
\end{equation}
The result \eqref{c12-approx} is also applicable for the light entropy modes with $\mn\ll1$. Substituting $\mu_1 = \sqrt{9/4-\beta} \approx 3/2$ in Eq. \eqref{c12-approx}, we find
\begin{eqnarray}\label{c12-R-light}
c_{1,2} \approx 
i^{1-\mu _2} e^{i x_c} \frac{
	\mu_2 \Gamma(\mu_2) }{\sqrt{2\pi }}
\left( \frac{x_{\rm c}^2 + 2 i x_{\rm c} - 2}{x_{\rm c}^{3/2}} \right)
\left(\frac{2}{y_{\rm c}}\right)^{\mu_2} \,,
\end{eqnarray}
which after substituting in \eqref{C-general0} gives
\begin{eqnarray}\label{C-light}
\Omega_{\tiny s,{\rm e}}^{\rm L}(\Delta{\cal N},p) &=& \frac{H_{\rm e}^2}{3 \Mp^2}
\frac{\mu_2^2 \Gamma(\mu_2)^2 }{8 \pi^3 }
\left(\frac{2}{y_{\rm c}}\right)^{2\mu_2}
\int_{x_{\rm min}}^{x_{\rm max}} \dd x\, 
\Big( L_{\rm e}
+ \left( \mr_{\rm e} +x^2 \right) K_{\rm e} \Big) 
\Big( x^3 e^{\Delta{\cal N}} + \frac{4}{x} e^{-3\Delta{\cal N}}\Big) 
\nonumber \\
&\approx& \frac{H_{\rm e}^2}{24 \pi^2 \Mp^2}
\frac{\Gamma(1/2p)^2 }{4 \pi p^2} \left(\frac{2p}{\sqrt{2}}\right)^{1/p} \Big( 
L_{\rm e} + \frac{10}{3} (1+e^{2p\Delta{\cal N}}) K_{\rm e} \Big) 
\big( 1+e^{2p\Delta{\cal N}} \big)^2 \,.
\end{eqnarray}
In obtaining the above result, we have neglected contributions from the lower limit defined in Eq. \eqref{tachyonic-light} in comparison with the dominant contributions from the upper limit. Using this result in Eq. \eqref{relic-DM0} we find the relic density for the dark matter produced by the light entropy modes as follows
\begin{eqnarray}\label{relic-light}
\Omega_{\tiny s,0}^{\rm L}(\Delta{\cal N},p,\mn_{\rm e}) =
{\cal O}(10^{20})
\mn_{\rm e}^{1/4} \Big(\frac{T_{\rm r}}{10^{12}{\rm GeV}}\Big)\, 
\Omega_{\tiny s,{\rm e}}^{\rm L}(\Delta{\cal N},p) \,.
\end{eqnarray} 

\section{Summary and Discussions}\label{Conclusion}

We have considered the most general two-field inflationary scenario with linear kinetic terms characterized by a curved field space and without higher derivative terms. Multiple field inflationary scenarios generally provide entropy perturbations which are constrained on CMB scales but otherwise may have interesting cosmological implications.  In the case of a flat field space, the superhorizon entropy modes are excited through the gravitational particle production process in a similar way as the curvature perturbations are stretched on superhorizon scales. The accumulated energy density of the excited superhorizon entropy modes can play the role of dark matter after the time of matter and radiation equality. These types of dark matter scenarios were already studied in the literature and it was found that only light entropy modes (compared with the Hubble expansion rate during inflation) can be excited through the gravitational instability. 

In this paper, we have looked at the role of the curvature of the field space. We have shown that even subhorizon heavy and semiheavy entropy modes (with the mass larger than or comparable to the Hubble expansion rate during inflation) can be excited through the tachyonic instability induced by the negative curvature of the field space which is known as the geometrical destabilization. We have obtained the spectral energy density, the spectral tilt, and the accumulated energy density of the excited entropy modes. The spectrum is blue-tilted so that the subhorizon modes provide the dominant contribution to the dark matter energy density. Compared to the models which are based on a flat field space and where only light superhorizon entropy modes are excited, the spectral energy density in our model has a peak at much smaller scales. Moreover, contrary to the vector dark matter models, our model does not provide any small scale vector-type anisotropies. These differences make our model observationally distinguishable from both vector dark matter models and scalar isocurvature dark matter scenarios based on the flat field space.

In order to simplify the analysis we have restricted our considerations to the geodesic motions in the field space corresponding to $\eta_\perp=0$. In general a non-geodesic motion can have interesting effects on the dark matter production in various ways. First, the parameter $\eta_\perp$ contribute negatively to the effective mass squared which can facilitate tachyonic instability (i.e. with $\beta <0$)
along with the negative field space curvature. Secondly, with $\eta_\perp\neq 0$ the adiabatic and the entropy modes are coupled at the linear order which can affect the production of the dark matter from the entropy modes and also the constraint on them from the CMB observations. We would like to come back to this question in future works.  

In addition, we have considered the simplified picture of an instant reheating.  In a realistic situation, during the (p)reheating process the scalar fields oscillate rapidly while the inflaton energy is transferred to SM particles. In a multiple field setup with a curved field space, this process is highly non-trivial and tachyonic instabilities can efficiently be activated during (p)reheating \cite{Krajewski:2018moi, Iarygina:2020dwe, Iarygina:2018kee}. This may generate small-scale entropy modes as the seed of the dark matter. In a sense what we have calculated in the current simplified setup with an instant reheating is a lower bound on the fractional density of dark matter from the entropy modes. It is an interesting question to look for the production of dark matter particles through the tachyonic resonance of entropy modes in a curved field space during (p)reheating.

%%%%%%%%%%%%%%%%%%%%%%%%%%%%%%%%%%%%%%%%%%%%%
\vspace{0.7cm}

{\bf Acknowledgments:} We thank Borna Salehian for insightful discussions and also sharing his code and Sébastien Renaux-Petel for helpful comments and discussions. M.A.G. thanks School of Astronomy at Institute for Research in Fundamental Sciences (IPM) for their hospitality during the final stage of this work. The work of M.A.G. was supported by Japan Society for the Promotion of Science Grants-in-Aid for international research fellow No. 19F19313. S.~M.'s work was supported in part by Japan Society for the Promotion of Science Grants-in-Aid for Scientific Research No.\ 17H02890, No.\ 17H06359, and by World Premier International Research Center Initiative, MEXT, Japan. H. F. and A. T. acknowledge partial support from the ``Saramadan" federation of Iran. 

\appendix

\section{Energy density for the entropy modes}\label{app-rho}

In this appendix, we find the energy density of the entropy modes. 

The energy-momentum tensor for our model defined by the action \eqref{total-action} is given by Eq. \eqref{EEs} as follows
\begin{eqnarray}\label{EMT}
T^{\mu}{}_{\nu} = -2 X^\mu{}_\nu + \big( X - V \big) \delta^\mu{}_\nu \,,
\end{eqnarray}
where we have defined
\begin{eqnarray}
X \equiv  g^{\mu\nu} X_{\mu\nu} = \gamma_{ab} X^{ab} \,,
\end{eqnarray}
with 

\begin{eqnarray}
X_{\mu\nu} \equiv 
- \frac{1}{2} \gamma_{a b}\partial_{\mu}\phi^a\partial_{\nu}\phi^b \,, 
\hspace{1cm} \mbox{and} \hspace{1cm}
X^{ab} \equiv 
- \frac{1}{2}g^{\alpha\beta} \partial_{\alpha}\phi^a\partial_{\beta}\phi^b \,.
\end{eqnarray}

The total energy density is then given by
\begin{eqnarray}\label{rho}
\rho = - T^0{}_0 = \gamma_{ab} \big( 2 {\bar X}^{ab} - X^{ab} \big) + V \,; \hspace{1cm}
{\bar X}^{ab} \equiv - \frac{1}{2} g^{\alpha {0}} \partial_\alpha\phi^a \dot{\phi}^b \,.
\end{eqnarray}
The energy density of the entropy modes is encoded in the second order perturbations of the energy density,
\begin{eqnarray}\label{rho-2}
\rho^{(2)} = \frac{1}{2} \Big[
\gamma_{ab}^{(2)} \Big( 2 {\bar X}^{ab(0)} - X^{ab(0)} \Big) 
+ \gamma_{ab}^{(0)} \Big( 2 {\bar X}^{ab(2)} - X^{ab(2)} \Big) + V^{(2)} 
\Big] \,,
\end{eqnarray}
where the upper indices $(i)$ show the order of perturbations. We work in spatially flat gauge by fixing the diffeomorphism gauge freedom so that $\psi=0=E$. Moreover, we focus on the decoupling limit and neglect metric perturbations $A$ and $B$ as their contributions are slow-roll suppressed. In this respect, the metric takes the unperturbed form of Eq. \eqref{metric}. Then, all information of the scalar perturbations are encoded in the scalar field perturbations $\delta\phi^a$ defined in Eq. \eqref{field-pert}. In this case, we find \cite{Gong:2011uw}
\begin{eqnarray}\label{X-pert}
{\bar X}^{ab(2)} &=& 
- g^{00(0)} \Big[ {\mathbb R}^{(a}{}_{ecd} \, \dot{\varphi}^{b)} \dot{\varphi}^{d}
\delta \phi^e \delta\phi^c 
+ \delta\dot{\phi}^a \delta\dot{\phi}^b \Big] \,, 
 \\
{X}^{ab(2)} &=& 
- g^{\mu\nu(0)} \Big[ {\mathbb R}^{(a}{}_{ecd} \, \partial_\mu\varphi^{b)} \partial_\nu{\varphi}^{d}
\delta \phi^e \delta\phi^c 
+ \partial_\mu \delta{\phi}^a \partial_\nu \delta{\phi}^b \Big] \,,
\nonumber \\ \nonumber
V^{(2)} &=& V_{ab} \, \delta\phi^a \delta\phi^b \,.
\end{eqnarray}

Note that $\gamma_{ab}^{(2)}$ identically vanishes through the metric compatibility condition with respect to the Christoffel symbols \eqref{christoffel} as explained in Ref. \cite{Gong:2011uw}. In the spatially flat gauge with $\psi=0$, from Eq. \eqref{curv-isocurv} we find
\begin{equation}\label{delta-phi}
\delta\phi^a = \frac{\dot{\sigma}}{H} {\cal R} \, T^a + {\cal F} \, N^a \,, \hspace{1cm}
\delta\dot{\phi}^a = {\Big(\frac{\dot{\sigma}}{H} {\cal R} \Big) }^{.}\,\, T^a + \dot{\cal F} N^a \,, 
\end{equation}
where we have used the fact that $\dot{T}^a = 0 = \dot{N}^a$ for $\eta_\perp=0$ from Eqs. \eqref{geodesics}. Substituting \eqref{delta-phi} in Eqs. \eqref{X-pert} and using the result in \eqref{rho-2} we find the following result for the energy density of the entropy modes
\begin{eqnarray}\label{app-rho-s}
\rho_s(\tau) = \frac{1}{2a^4} \bigg[\,
\Big(a\left(\dfrac{s}{a}\right)'\Big)^2 + (\nabla{s})^2
+ \big( V_{NN} - \mR^2 \big) a^2 s^2
\bigg] \,,
\end{eqnarray}
where we have worked with the conformal time and with the canonically normalized entropy field $s=a{\cal F}$. In obtaining the above result we have also used Eqs. \eqref{Riemann}, \eqref{T-N}, and \eqref{T-N-norm}. Going to the Fourier space, we find the result \eqref{rho-s} for the energy density of the entropy modes at the end of inflation $\tau=\tau_{\rm e}$.

\section{Hamiltonian for the entropy modes}\label{app-Hamiltonian}

In this appendix we compute the quadratic Hamiltonian for the entropy modes based on the quadratic action \eqref{conformal-action-R-F}.

As the curvature and entropy perturbations completely decouple from each other at the linear order for the geodesic trajectory $\eta_{\perp}=0$, we rewrite the action \eqref{conformal-action-R-F} as follows
\begin{eqnarray}\label{app-action-tot}
S^{(2)}%_{\rm tot} 
=  S^{(2)}_{\cal R} + S^{(2)}_{\cal F} \,,
\end{eqnarray}
where we have defined
\begin{eqnarray}\label{app-action-entropy}
S^{(2)}_{\cal F} \equiv \int \! \, {\rm d}\tau \, {\cal L}^{(2)}_{\cal F} \,;
\hspace{1cm}
{\cal L}^{(2)}_{\cal F} \equiv  \frac{a^2}{2} \int {\rm d}^3 x \Big[ {\cal F}^{'2}  
- (\nabla \mathcal F)^2 - \big( V_{NN} + \mR^2 \big) a^2 \mathcal F^2 \Big] \,.
\end{eqnarray}

In \eqref{app-action-tot}, we have also defined $S^{(2)}_{\cal R} \equiv S^{(2)}%_{\rm tot}
-S^{(2)}_{\cal F}$ in which $S^{(2)}_{\cal R}$ is the quadratic action for the curvature perturbations in the case of $\eta_{\perp}=0$. Defining the canonical momentum $\pi^{(2)}_{\cal F} = \partial{\cal L}^{(2)}_{\cal F}/\partial {\cal F}'$ and performing the Legendre transformation, the Hamiltonian for the entropy modes turns out to be
\begin{eqnarray}\label{app-Hamiltonian-F}
{\cal H}^{(2)}_{\cal F} = {\cal F}' \pi^{(2)}_{\cal F} - {\cal L}^{(2)}_{\cal F}
 =  \frac{a^2}{2} \int {\rm d}^3 x \Big[ {\cal F}^{'2}  
 + (\nabla \mathcal F)^2 + \big( V_{NN} + \mR^2 \big) a^2 \mathcal F^2 \Big]
\,.
\end{eqnarray}

Going to the Fourier space and working with the canonical entropy field $s=a{\cal F}$, which is defined in Eq. \eqref{canonical-fields}, we find
\begin{eqnarray}\label{app-Hamiltonian-s}
{\cal H}^{(2)}_{s} =  \frac{1}{2} \int
\dfrac{\dd^3 k}{(2\pi)^3}  \, \bigg[\,
\Big|
a\left(\dfrac{s_k}{a}\right)'
\Big|^2  + 
\Big(
k^2+  \big( V_{NN} + \mR^2 \big) a^2
\Big)
|s_k|^2
\bigg] \,.
\end{eqnarray}

If we seek the energy density associated with the above Hamiltonian through the relation ${\cal H}^{(2)}_s/\sqrt{-g}={\cal H}^{(2)}_s/a^4$, we find a result different from Eq. \eqref{rho-s} that we find from the second order expansion of the energy-momentum tensor Eq. \eqref{EMT}. More precisely, the sign of the mass term induced by the curvature of the field space $\mR^2$ is different for these two energy densities. First, this is not surprising as the second order part of the energy-momentum tensor reflects a part of the cubic action while the Hamiltonian \eqref{app-Hamiltonian-s} is the generator of the time evolution for the linear perturbations which is obtained from the quadratic action. As is known, these quantities are different in general and this is the case in our model. Secondly, the different sign for $\mR^2$ term in Eqs. \eqref{app-rho-s} and \eqref{app-Hamiltonian-s} shows that $\rho_s>0$ and ${\cal H}^{(2)}_{s}<0$ for $\mR^2<0$ ($\mr>0$). Thus, the system becomes unstable in this regime which is a signature of the existence of tachyonic instability in our model. Indeed, we need this local instability induced by the negative curvature of the field space to produce entropy modes in the context of the so-called geometrical destabilization mechanism \cite{Renaux-Petel:2015mga}. 

%=====================================================================

%\bibliographystyle{JHEP}
\bibliography{references}
%\bibliography{Bibs}{} 
\end{document}